%Paper: hep-th/9403192
%From: cvj@guinness.ias.edu (Clifford Johnson)
%Date: Thu, 31 Mar 94 10:10:59 EST
%Date (revised): Thu, 31 Mar 94 10:24:10 EST
%Date (revised): Thu, 31 Mar 94 10:39:51 EST
%Date (revised): Thu, 31 Mar 94 10:53:16 EST
%Date (revised): Sun, 26 Jun 94 16:23:23 EDT

%%%%%%%%%%%%%%%%  preprint and letter macro  %%%%%%%%%%%%%%%%%

\hsize=6.0truein
\vsize=8.5truein
\voffset=0.25truein
\hoffset=0.1875truein%would be 0.25 except our laser printer is off by 1/16in
\tolerance=1000
\hyphenpenalty=500
\def\monthintext{\ifcase\month\or January\or February\or
   March\or April\or May\or June\or July\or August\or
   September\or October\or November\or December\fi}

%%%%%%%%%%%%%%%%%  Twelve point text font  %%%%%%%%%%%%%%%%%%%

\font\tenrm=cmr10 scaled \magstep1   \font\tenbf=cmbx10 scaled \magstep1
\font\sevenrm=cmr7 scaled \magstep1  
\font\fiverm=cmr5 scaled \magstep1   

\font\teni=cmmi10 scaled \magstep1   \font\tensy=cmsy10 scaled \magstep1
\font\seveni=cmmi7 scaled \magstep1  \font\sevensy=cmsy7 scaled \magstep1
\font\fivei=cmmi5 scaled \magstep1   \font\fivesy=cmsy5 scaled \magstep1

\font\tentt=cmtt10 scaled \magstep1
\font\tenit=cmti10 scaled \magstep1
\font\tensl=cmsl10 scaled \magstep1

\def\twelvepoint{\def\rm{\fam0\tenrm}
   \textfont0=\tenrm \scriptfont0=\sevenrm \scriptscriptfont0=\fiverm
   \textfont1=\teni  \scriptfont1=\seveni  \scriptscriptfont1=\fivei
   \textfont2=\tensy \scriptfont2=\sevensy \scriptscriptfont2=\fivesy
   \textfont\itfam=\tenit \def\it{\fam\itfam\tenit}
   \textfont\ttfam=\tentt \def\tt{\fam\ttfam\tentt}
   \textfont\bffam=\tenbf \def\bf{\fam\bffam\tenbf}
   \textfont\slfam=\tensl \def\sl{\fam\slfam\tensl} \rm
   %Essentially I changed all dimensions to 1.2 times as large as in plain tex
   \hfuzz=1pt\vfuzz=1pt%much more than plain tex's value
   \setbox\strutbox=\hbox{\vrule height 10.2pt depth 4.2pt width 0pt}
   \parindent=24pt\parskip=1.2pt plus 1.2pt
   \topskip=12pt\maxdepth=4.8pt\jot=3.6pt
   \normalbaselineskip=14.4pt\normallineskip=1.2pt
   \normallineskiplimit=0pt\normalbaselines
   \abovedisplayskip=13pt plus 3.6pt minus 5.8pt
   \belowdisplayskip=13pt plus 3.6pt minus 5.8pt
   \abovedisplayshortskip=-1.4pt plus 3.6pt
   \belowdisplayshortskip=13pt plus 3.6pt minus 3.6pt
   %plain tex's value for belowdisplayshortskip looked terrible
   \topskip=12pt \splittopskip=12pt
   \scriptspace=0.6pt\nulldelimiterspace=1.44pt\delimitershortfall=6pt
   \thinmuskip=3.6mu\medmuskip=3.6mu plus 1.2mu minus 1.2mu
   \thickmuskip=4mu plus 2mu minus 1mu%reduced these plain tex values
   \smallskipamount=3.6pt plus 1.2pt minus 1.2pt
   \medskipamount=7.2pt plus 2.4pt minus 2.4pt
   \bigskipamount=14.4pt plus 4.8pt minus 4.8pt}

\twelvepoint

%%%%%%%%%%%%%%%%% Definitions for Preprints %%%%%%%%%%%%%%%%%%

% title page title font

\font\titlerm=cmr10 scaled \magstep3
\font\titlerms=cmr10 scaled \magstep1 %\font\titlermss=cmr8
\font\titlei=cmmi10 scaled \magstep3  %math italic for title
\font\titleis=cmmi10 scaled \magstep1 %\font\titleiss=cmmi8
\font\titlesy=cmsy10 scaled \magstep3 	%math symbols for title
\font\titlesys=cmsy10 scaled \magstep1  %\font\titlesyss=cmsy8
\font\titleit=cmti10 scaled \magstep3	%text italic for title
\skewchar\titlei='177 \skewchar\titleis='177 %\skewchar\titleiss='177
\skewchar\titlesy='60 \skewchar\titlesys='60 %\skewchar\titlesyss='60

\def\titlefont{\def\rm{\fam0\titlerm}% switch to title font
   \textfont0=\titlerm \scriptfont0=\titlerms %\scriptscriptfont0=\titlermss
   \textfont1=\titlei  \scriptfont1=\titleis  %\scriptscriptfont1=\titleiss
   \textfont2=\titlesy \scriptfont2=\titlesys %\scriptscriptfont2=\titlesyss
   \textfont\itfam=\titleit \def\it{\fam\itfam\titleit} \rm}

% title page macros

\def\preprint#1{\baselineskip=19pt plus 0.2pt minus 0.2pt \pageno=0
   \begingroup%use with \draft or \date to end group
   \nopagenumbers\parindent=0pt\baselineskip=14.4pt\rightline{#1}}
\def\title#1{
   \vskip 0.9in plus 0.45in
   \centerline{\titlefont #1}}
\def\secondtitle#1{}%set up this some time
\def\author#1#2#3{\vskip 0.9in plus 0.45in
   \centerline{{\bf #1}\myfoot{#2}{#3}}\vskip 0.12in plus 0.02in}
\def\secondauthor#1#2#3{}%set up this some time
\def\addressline#1{\centerline{#1}}
\def\abstract{\vskip 0.7in plus 0.35in
	\centerline{\bf Abstract}
	\smallskip}
\def\finishtitlepage#1{\vskip 0.8in plus 0.4in
   \leftline{#1}\supereject\endgroup}

\def\date#1{\finishtitlepage{#1}}

\def\nolabels{\def\eqnlabel##1{}\def\eqlabel##1{}\def\figlabel##1{}%
	\def\reflabel##1{}}
\def\writelabels{\def\eqnlabel##1{%
	{\escapechar=` \hfill\rlap{\hskip.11in\string##1}}}%
	\def\eqlabel##1{{\escapechar=` \rlap{\hskip.11in\string##1}}}%
	\def\figlabel##1{\noexpand\llap{\string\string\string##1\hskip.66in}}%
	\def\reflabel##1{\noexpand\llap{\string\string\string##1\hskip.37in}}}
\nolabels

%  tagged section numbers

\global\newcount\secno \global\secno=0
\global\newcount\meqno \global\meqno=1
\global\newcount\subsecno \global\subsecno=0

\font\secfont=cmbx12 scaled\magstep1

\def\section#1{\global\advance\secno by1
   \xdef\secsym{\the\secno.}
   \global\subsecno=0
   \global\meqno=1\bigbreak\medskip
   \noindent{\secfont\the\secno. #1}\par\nobreak\smallskip\nobreak\noindent}
%\xdef\secsym{}

\def\subsection#1{\global\advance\subsecno by1
    %\xdef\secsym{\the\subsecno}
\medskip
\noindent
{\bf\the\secno.\the\subsecno\ #1}
\par\medskip\nobreak\noindent}
%\xdef\secsym{}

\def\newsec#1{\global\advance\secno by1
   \xdef\secsym{\the\secno.}
   \global\meqno=1\bigbreak\medskip
   \noindent{\bf\the\secno. #1}\par\nobreak\smallskip\nobreak\noindent}
\xdef\secsym{}

\def\appendix#1#2{\global\meqno=1\xdef\secsym{\hbox{#1.}}\bigbreak\medskip
\noindent{\bf Appendix #1. #2}\par\nobreak\smallskip\nobreak\noindent}

%         equations

\def\eqnn#1{\xdef #1{(\secsym\the\meqno)}%
	\global\advance\meqno by1\eqnlabel#1}
\def\eqna#1{\xdef #1##1{\hbox{$(\secsym\the\meqno##1)$}}%
	\global\advance\meqno by1\eqnlabel{#1$\{\}$}}
\def\eqn#1#2{\xdef #1{(\secsym\the\meqno)}\global\advance\meqno by1%
	$$#2\eqno#1\eqlabel#1$$}

%			 footnotes

\def\myfoot#1#2{{\baselineskip=14.4pt plus 0.3pt\footnote{#1}{#2}}}
%sequentially numbered footnotes
\global\newcount\ftno \global\ftno=1
\def\foot#1{{\baselineskip=14.4pt plus 0.3pt\footnote{$^{\the\ftno}$}{#1}}%
	\global\advance\ftno by1}

%         references

\global\newcount\refno \global\refno=1
\newwrite\rfile

\def\ref{[\the\refno]\nref}
\def\nref#1{\xdef#1{[\the\refno]}\ifnum\refno=1\immediate
	\openout\rfile=refs.tmp\fi\global\advance\refno by1\chardef\wfile=\rfile
	\immediate\write\rfile{\noexpand\item{#1\ }\reflabel{#1}\pctsign}\findarg}
%	horrible hack to sidestep tex \write limitation
\def\findarg#1#{\begingroup\obeylines\newlinechar=`\^^M\passarg}
	{\obeylines\gdef\passarg#1{\writeline\relax #1^^M\hbox{}^^M}%
	\gdef\writeline#1^^M{\expandafter\toks0\expandafter{\striprelax #1}%
	\edef\next{\the\toks0}\ifx\next\null\let\next=\endgroup\else\ifx\next\empty%

\else\immediate\write\wfile{\the\toks0}\fi\let\next=\writeline\fi\next\relax}}
	{\catcode`\%=12\xdef\pctsign{%}}
\def\striprelax#1{}

\def\semi{;\hfil\break}
\def\addref#1{\immediate\write\rfile{\noexpand\item{}#1}} %now unnecessary

\def\listrefs{\vfill\eject\immediate\closeout\rfile
   {{\secfont References}}\bigskip{\frenchspacing%
   \catcode`\@=11\escapechar=` %
   \input refs.tmp\vfill\eject}\nonfrenchspacing}

\def\startrefs#1{\immediate\openout\rfile=refs.tmp\refno=#1}

%		and finally, figures:

\global\newcount\figno \global\figno=1
\newwrite\ffile
\def\fig{\the\figno\nfig}
\def\nfig#1{\xdef#1{\the\figno}\ifnum\figno=1\immediate
	\openout\ffile=figs.tmp\fi\global\advance\figno by1\chardef\wfile=\ffile
	\immediate\write\ffile{\medskip\noexpand\item{Fig.\ #1:\ }%
	\figlabel{#1}\pctsign}\findarg}

\def\listfigs{\vfill\eject\immediate\closeout\ffile{\parindent48pt
	\baselineskip16.8pt{{\secfont Figure Captions}}\medskip
	\escapechar=` \input figs.tmp\vfill\eject}}

%%%%%%%%%%%%%%%%%%%%%%%%%%%%%%%%%%%%%%%%%%%%%%%%%%%%%%%%%%%%%%%%%%%%%%%%%%%%%%%
\def\noblackbox{\overfullrule=0pt}
\def\inv{^{\raise.18ex\hbox{${\scriptscriptstyle -}$}\kern-.06em 1}}
\def\dup{^{\vphantom{1}}}
\def\Dsl{\,\raise.18ex\hbox{/}\mkern-16.2mu D} %this one can be subscripted
\def\dsl{\raise.18ex\hbox{/}\kern-.68em\partial}
\def\slash#1{\raise.18ex\hbox{/}\kern-.68em #1}
\def\lspace{}
\def\lbspace{}
\def\boxeqn#1{\vcenter{\vbox{\hrule\hbox{\vrule\kern3.6pt\vbox{\kern3.6pt
	\hbox{${\displaystyle #1}$}\kern3.6pt}\kern3.6pt\vrule}\hrule}}}
\def\mbox#1#2{\vcenter{\hrule \hbox{\vrule height#2.4in
	\kern#1.2in \vrule} \hrule}}  %e.g. \mbox{.1}{.1}
%matters of taste
%\def\tilde{\widetilde}
\def\bar{\overline}
\def\e#1{{\rm e}^{\textstyle#1}}
\def\del{\partial}
\def\curly#1{{\hbox{{$\cal #1$}}}}
\def\curlyD{\hbox{{$\cal D$}}}
\def\curlyL{\hbox{{$\cal L$}}}
\def\vev#1{\langle #1 \rangle}
\def\psibar{\overline\psi}
\def\lform{\hbox{$\sqcup$}\llap{\hbox{$\sqcap$}}}
\def\darr#1{\raise1.8ex\hbox{$\leftrightarrow$}\mkern-19.8mu #1}
\def\half{{\textstyle{1\over2}}} %puts a small half in a displayed eqn
\def\roughly#1{\ \lower1.5ex\hbox{$\sim$}\mkern-22.8mu #1\,}
\def\MSbar{$\bar{{\rm MS}}$}
%%%%%%%%%%%%%%%%%%%%%%%%%%%%%%%%%%%%%%%%%%%%%%%%%%%%%%%%%%%%%%
\hyphenation{di-men-sion di-men-sion-al di-men-sion-al-ly}

\parindent=0pt
\parskip=5pt

\def\Tr{{\rm Tr}}
\def\det{{\rm det}}
\def\jump{\hskip1.0cm}
\def\wzw{Wess--Zumino--Witten}
\def\Az{A_z}
\def\Azb{A_{\bar{z}}}
\def\lr{\lambda_R}
\def\ll{\lambda_L}
\def\lrb{\bar{\lambda}_R}
\def\llb{\bar{\lambda}_L}
\font\top = cmbxti10 scaled \magstep1

\def\d{\partial_z}
\def\db{\partial_{\bar{z}}}
\def\rline{{{\rm I}\!{\rm R}}}
\def\tl{t_L}
\def\tr{t_R}

%%%%%%%%%%%%%%%%%%%%%%%%%%%%%%% PAPER STARTS HERE %%%%%%%%%%%%%%%%%%%%%%%%%%%
\preprint{
\vbox{
\rightline{IASSNS--HEP--94/20}
\vskip2pt\rightline{hep-th/9403192}
\vskip2pt\rightline{March 1994.}
\vskip2pt\rightline{(Revised June 1994).}
}
}
\vskip -1cm

\title{
\vbox{
\centerline{ Exact Models of }
\vskip2pt\centerline{Extremal Dyonic 4D Black Hole Solutions}
\vskip2pt\centerline{of Heterotic String Theory.}
}
}
\vskip -1.5cm
\author{\bf Clifford V Johnson\myfoot{$^*$}{\rm  SERC (UK) Fellow. Part of this
work was supported by a Lindemann Fellowship.}\myfoot{$^\dagger$}{\rm e-mail:
cvj@guinness.ias.edu}}{}{}
\vskip 0.7cm
\addressline{\it School of Natural Sciences}
\addressline{\it Institute for Advanced Study}
\addressline{\it Olden Lane}
\addressline{\it Princeton, NJ 08540 U.S.A.}
\vskip -1.2cm

\abstract

{
Families of exact $(0,2)$ supersymmetric conformal field theories of
magnetically and electrically charged  extremal 4D black hole  solutions of
heterotic string theory are presented. They are constructed  using a $(0,1)$
supersymmetric  $SL(2,\rline)\times SU(2)$ \wzw\ model where  anomalously
embedded $U(1)\times U(1)$ subgroups are gauged. Crucial  cancelations of the
$U(1)$ anomalies coming from  the supersymmetric fermions, the current algebra
fermions and the gauging ensure that there is a consistency of these models at
the quantum level. Various 2D models, which may be considered as building
blocks for extremal 4D constructions, are presented. They generalise the class
of 2D models which might be obtained from gauging $SL(2,\rline)$ and coincide
with known heterotic string backgrounds. The exact conformal field theory
presented by Giddings, Polchinski and Strominger describing the angular sector
of the extremal magnetically charged black hole  is a special case of  this
construction. An example where the radial and angular theories are mixed
non--trivially is studied in detail, resulting in an extremal dilatonic
Taub--NUT--like dyon.}

\vskip -1cm
\date{}
%\draft{}

\newsec{Introduction.}
Since Witten's observation\ref\edone{E Witten, Phys Rev {\bf D44} (1991) 314.}
that an exact conformal field theory of a 2D black hole background for string
theory may be obtained by an $SL(2,\rline)/U(1)$ gauged \wzw\  (WZW) model,
there has been much activity in the field  of string theory in non--compact
curved spacetime backgrounds   employing  similar techniques.

Due to the fact that non--compact Lie Groups in general possess continuous
spectra of unitary states, the conformal field theories derived from the WZW
models based on these groups is poorly understood at present. So although it
would be desirable to study the black hole and related conformal field theories
in as much detail as the more familiar conformal field theories based on
compact groups, little progress has been made beyond calculations which
severely truncate the spectra\foot{See for example refs.\ref\verlinde{R
Dijkgraaf, E Verlinde and H Verlinde, Nucl Phys {\bf B371} (1992)
269--314}\ref\samsoni{E J Martinec and  S L Shatashvili, Nucl Phys {\bf B368}
(1992) 338.}\ref\Lykken{S Chaudhuri and J D Lykken, Nucl Phys {\bf B396} (1993)
270. hep-th/9206107.} for some direct work on the black hole coset using
conformal field theory techniques.}.

While waiting for the development of the tools needed to calculate extensively
in these theories, researchers in the field have not been idle. There has been
much energy directed towards the  problem of finding more of this type of exact
solution. At the level of a sigma model the solutions are the all orders in
$\alpha^\prime$ solution to the $\beta$--function equations of the background
field problem in string theory\ref\callan{C G Callan, D Friedan, E J Martinec,
M J Perry,  Nucl Phys {\bf B262} (1985) 593--609.}. Beyond the $\alpha^\prime$
expansion, the non--perturbative information contained in these models describe
aspects of quantum gravity which may not be contained in a metric theory of
gravity.

Of particular interest is the problem of finding the conformal field theories
corresponding to 4D black hole solutions of string theory. This is because of
the expectation that string theory, as a possibly relevant quantum theory of
gravity, might contribute significantly novel and useful features to the
physics of these interesting objects. Two of the most pertinent problems here
are the black hole information paradox and the nature of physics at curvature
singularities. The exact conformal field theories
corresponding to these backgrounds contain, in principle, the answers that
string theory has to give about these problems.

 One of the first lessons from string theory in the classroom of black hole
physics is that the necessary presence of  scalar fields significantly
distinguishes  the physics of   stringy black holes from that of their
Einsteinian cousins.
Consider the case  when the black holes are charged.
The relevant scalar fields are the dilaton and modulus fields  arising from the
compactification of the six internal dimensions of the heterotic string
theory\foot{The author is  grateful to M J Duff for pointing out the role of
the modulus field. See ref.\ref\mike{M J Duff, R R Khuri, R Minasian and J
Rahmfeld, Nucl  Phys {\bf B418}, (1994) 195. hep-th/9311120.} for examples of
solutions of heterotic string theory which include modulus fields arising in
toroidal compactifications. }. They combine to give an effective
scalar--Maxwell coupling $\e{-\alpha\phi}F^2$ in the effective 4D theory, where
$\alpha\ge0$. In particular for $\alpha$ non--zero, the
  Reissner--Nordstr\o m solution, the prototype of the charged black hole in
General Relativity, is inapplicable as a solution  even in regions
of spacetime where the curvature is mild.  (This is in contrast to the
Schwarzchild solution, which together with a constant dilaton is a stringy
solution at one loop order in $\alpha^\prime$, and hence outside sufficiently
large neutral black holes.)

In this paper only the pure dilaton solutions will be
considered\foot{Ref.\ref\holzhey{C F E Holzhey and F Wilczek, Nucl Phys {\bf
B380} (1992) 447. hep-th/9202014.} includes an extensive discussion of
scalar--Maxwell black hole physics for more general values of $\alpha$.}, for
which $\alpha=1$. For this case, the model charged solution is not
Reissner--Nordstr\o m solution but instead the magnetically charged dilatonic
black hole
solution\ref\gibbons{G W Gibbons and K Maeda, Nucl Phys {\bf B298} (1988)
741\semi D Garfinkle, G T Horowitz and A Strominger Phys Rev {\bf D43} (1991)
3140, erratum Phys Rev {\bf D45} (1992) 3888.} and its dyonic generalisations,
 some of which were found by exploiting  an $SL(2,\rline)$ duality symmetry of
the equations of motion\ref\frank{A Shapere, S Trivedi and F Wilczek, Mod Phys
Lett {\bf A6} (1991) 2677.}.
These types of solutions were only known  to one loop order in $\alpha^\prime$
until relatively recently.
An exact solution for the spherically symmetric magnetically charged 4D black
hole solution to heterotic string theory was presented by Giddings, Polchinski
and Strominger\ref\GPS{S B Giddings, J Polchinski and A Strominger, Phys Rev
{\bf D48} (1993) 5784. hep-th/9305083.}\ (GPS).
This solution is the exact conformal field theory for which the dilatonic
magnetic black hole of Gibbons and Maeda\gibbons\ is the low energy limit.
Stated more precisely, it describes the extremal black hole   limit with an
infinite throat.
 The conformal field theory takes the form of a product of the supersymmetric
2D linear dilaton black hole for the radial $(\sigma,t)$ sector and a monopole
conformal field theory for the angular $(\theta,\phi)$ sector. The   linear
dilaton black hole is constructed  using a   supersymmetric gauged
$SL(2,\rline)/U(1)$   WZW model, while the monopole theory was identified as a
$Z_{2Q+2}$ orbifold of an $SU(2)$ WZW model. This construction will be briefly
reviewed in the next section.

The significance of this factorisation into a product form\foot{The process of
approaching extremality for the dilatonic magnetic black hole is described in
ref.\GPS\ and also discussed   in section 9.}\ is important. The physics of
stringy black hole evaporation (and other processes) studied in 2D is quite
relevant to   4D stringy black holes in the case of exact extremality. It is
presumably possible to expand about this point in parameter space  to study the
physics of near extremal black holes. Generically the theory decomposes into  a
`black hole $+$ wormhole' situation, where the black hole is at the bottom of a
semi--infinite wormhole. The wormhole has topology $\rline^2\times S^2$ in the
infinite `throat' region, and is asymptotic to Minkowski space after emerging
from the `mouth' region. In the throat region the two--sphere has a constant
radius set by the charge of the black hole. The conformal field theory of GPS
describes the black hole together with the semi--infinite throat. The same will
be the case for the solutions described in this paper. Little is known about
the conformal field theory describing the widening  transition from the throat
region  to the mouth of the wormhole. However operators describing marginal
perturbations of the conformal field theory, which represent  clearing  the
throat region have been identified and discussed\ref\fivebrane{S B Giddings and
A  Strominger, Phys Rev Lett, {\bf 67} (1991) 2930.}\GPS.

 There have also been some conformal field theories of bosonic strings in
axisymmetric 4D black hole backgrounds presented in refs.\ref\gershon{D
Gershon, {\sl `Exact solutions of 4D black holes in string theory'}, preprint
TAUP--2033--92, hep-th/9202005\semi D Gershon, Phys Rev {\bf D49} (1994) 999.
hep-th/9210160\semi D Gershon, {\sl `Semiclassical vs exact solutions of
charged black holes in four dimensions and exact $O(d,d)$ duality'}, preprint
TAUP--2121--93, hep-th/9311122.}. These solutions were obtained by gauging
anomaly--free subgroups of  $SL(2,\rline)\times SU(2)\times U(1)$ WZW models,
or equivalently by doing some $O(d,d)$ duality transformations on ungauged WZW
models.

The group $SL(2,\rline)\times SU(2)$ has played a prominent role in many of the
  constructions generalising Witten's work, starting with the rotating
solutions of ref.\ref\petr{P Horava, Phys Lett {\bf B278} (1992) 101.
hep-th/9110067.}\ and including a  cosmological solution\ref\chiaraEd{C R Nappi
and E Witten, Phys Lett {\bf B293} (1992) 309. hep-th/9206078}, with numerous
other constructions in the literature. This type of WZW model will also play a
central role in this paper, with many novel features. A family of exact
heterotic string solutions will be constructed as a $(0,1)$
supersymmetric\foot{The final model later has $(0,2)$ supersymmetry due to the
nature of the coset. See later sections for a discussion of this.}\
$SL(2,\rline)\times SU(2)$ WZW model with $U(1)\times U(1)$ subgroups gauged.
The subgroup will be embedded in a way which will produce a classical  anomaly
in contrast to the non--anomalous embeddings more frequently used. This
anomaly, together with the anomaly from the right--moving supersymmetric
fermions, will be cancelled against an anomaly coming from left--moving
fermions which have been put in by hand to play the role of the current algebra
fermions of the heterotic string theory. After integrating out the gauge
fields, the resulting heterotic sigma model possesses a spacetime
interpretation as extremal magnetically and electrically charged backgrounds
with a non--trivial metric, dilaton, gauge and antisymmetric tensor fields. The
metrics possesses  asymptotic throat regions.

The model of GPS may be understood as  a special case of the construction of
this paper\foot{This construction   was suggested to the author  by E Witten.},
and although it was not presented in terms of a gauged WZW with fermions, this
is a most natural framework within which to describe that model. This framework
readily leads to fruitful generalisations, to which end this paper is a first
step. The paper begins with a brief review in section 2 of the identification
by GPS of their angular theory together with world sheet fermions as a WZW
orbifold. Section 3 then recasts the theory as a gauged WZW model  as a
prototype illustration of the construction. It ends by discussing the
non--trivial task  of determining the  correct spacetime metric from the final
theory. This procedure differs from that which is involved in the more
traditional gaugings of WZW models because of the non--neglible effect of the
fermions on the spacetime. Section 4 constructs a large family of conformal
field theories based on $SL(2,\rline)$ with heterotic fermions using the basic
technique of this paper. The family includes the charged 2D black hole
solutions already known to one loop as discussed in section 5. Extremal 4D
solutions can then be constructed using products of the $SL(2,\rline)$ theories
with the $SU(2)$ theories.  The rest of the paper is devoted to constructing an
extremal dyonic 4D  solution as starting example of the rich class of heterotic
string backgrounds which might be constructed in the manner outlined, by mixing
the radial and angular theories non--trivially.

\vfill\eject
\newsec{The GPS Monopole theory.}

The identification of the monopole theory \'a la Giddings, Polchinski and
Strominger (GPS) begins with the $SU(2)$ WZW action at level $k$:
\eqn\Wzw{I_{WZW_k}=-{k\over4\pi}\int_\Sigma d^2z\,\,\,\Tr(g^{-1}\d g\cdot
g^{-1}\db g)-
ik\Gamma(g)
}
where
\eqn\gam{\Gamma(g)=\int_B g^*\omega.
}

As is familiar this model is a theory of maps $g:\Sigma\to SU(2)$. $\Sigma$ is
($1+1$ dimensional) spacetime, the boundary of a ball $B$.
Here $\omega$ is a $SU(2)_L\times SU(2)_R$ invariant 3--form on the manifold of
$SU(2)$, normalised so that its integral over the whole manifold  is $2\pi$.

It is convenient to use the following Euler parameterisation for $g\in SU(2)$:
\eqn\euler{g=\e{i\phi\sigma_3/2}\e{i\theta\sigma_2/2}\e{i\psi\sigma_3/2}
=\pmatrix{
\e{{i\over2}\phi_+}\cos{\theta\over2}&
\e{{i\over2}\phi_-}\sin{\theta\over2}\cr
-\e{-{i\over2}\phi_-}\sin{\theta\over2}&
\e{-{i\over2}\phi_+}\cos{\theta\over2}\cr
},}
where the $\sigma_i$ are the Pauli matrices and
\eqn\ranges{\phi_\pm\equiv\phi\pm\psi,\jump
0\leq\theta\leq\pi,\jump0\leq\phi\leq2\pi,\jump0\leq\psi\leq4\pi.}
This gives
\eqn\sthree{\eqalign{I_{WZW_k}&={k\over8\pi}\int d^2z
\,\,\,\left\{G^{S_3}_{\alpha\beta}\d X^\alpha\db X^\beta\right\}-i\Gamma(g)=\cr
&={k\over8\pi}\int d^2z\,\,\,
\left\{\d\theta\db\theta+\d\phi\db\phi
+\d\psi\db\psi+\cos\theta(\d\phi\db\psi+\d\psi\db\phi)\right\}\cr
&+{k\over8\pi}\int d^2z\,\,\,(\pm1-\cos\theta)(\db\phi\d\psi-\db\psi\d\phi).
}}

The Wess--Zumino term is constructed using the fact that the unique (up to
scalings) $G_L\times G_R$ invariant three--form on $SU(2)$ is the volume form
$\omega_V$ of the three--sphere $S_3$, which is $\omega_V=|\det
G^{S_3}|^{1/2}d\theta\wedge d\phi\wedge d\psi$. The metric is
\eqn\metric{G=\pmatrix{1&0&0\cr
0&1&\cos\theta\cr0&\cos\theta&1},} and therefore (given that the volume of
$S_3$ is $4\pi^2$)
\eqn\omeg{\omega={1\over8\pi}\sin\theta d\theta\wedge d\phi\wedge d\psi.}

The Wess--Zumino term is written as a two--dimensional integral by  writing
$\omega=d\lambda$ locally. This has solution
$\lambda={1\over8\pi}(\pm1-\cos\theta)d\phi\wedge d\psi$ where the $\pm$ choice
corresponds to the North and South poles $\theta=0,\pi$ respectively. Using
this together with $\epsilon^{z\bar{z}}=i=-\epsilon^{\bar{z}z}$ yields
\eqn\wesszum{\eqalign{
\Gamma(g)=\int_Bg^*\omega=\int_\Sigma g^*\lambda=&
\int d^2z\,\,\epsilon^{ab}\lambda_{ij}\partial_a\phi^i\partial_b\psi^j\cr
=&{i\over8\pi}\int d^2z\,\,(\pm1-\cos\theta)(\db\phi\d\psi-\db\psi\d\phi)
}
}
 Now note that the $S^3$ metric may be decomposed as follows:
$$\eqalign{
G^{S_3}_{\alpha\beta}\d X^\alpha \db X^\beta&=(G^{S_2}_{\mu\nu}+4A^M_\mu
A^M_\nu)\d X^\mu \db X^\nu\cr &-(1\mp2\cos\theta)\d\phi\db\phi+
\d\psi\db\psi+\cos\theta(\d\phi\db\psi+\d\psi\db\phi).}
$$
Here $X^1=\theta$ and $X^2=\phi$ and  $A_\phi^M$
is the    non--zero component of the field $A^M_\mu$ of a magnetic monopole:
\eqn\monfield{A^M=\pmatrix{0\cr{\pm1-\cos\theta\over2}},} where the `$\pm$'
choice refers to the Northern or Southern hemisphere of $S^2$.

In \GPS\ the identification of the heterotic sigma model is made with a \wzw\
model by  identifying   $\psi=\xi\mp\phi$ in the North (South) pole. Then with
$\xi=X^3/Q_+$, and setting the  level $k=2Q_+Q_-$,   the model is identical to
that defined in equation (3.14) of Giddings, Polchinski and Strominger:
\eqn\gps{\eqalign{I_{GPS}={Q_+Q_-\over4\pi}\int d^2z\,\,\,\Biggl\{
(G^{S_2}_{\mu\nu}&+4A^M_\mu A^M_\nu)\d X^\mu \db X^\nu\cr
&+{1\over Q_+^2}(\d X^3-4Q_+A^M_\mu\d X^\mu)\db X^3\Biggr\}.}}
(Note that $Q_\pm\equiv Q\pm1$)

 In \GPS\ the bosonic field $X^3$ arose in the sigma model by bosonising world
sheet fermions and is $2\pi$ periodic. In order to complete the identification
therefore, given that $\xi$ is a $4\pi$ periodic field the further
identification $\xi\sim\xi+2\pi/Q_+$ must be made, and therefore  the sigma
model of \GPS\ was identified as
 an $SU(2)/Z_{2Q_+}$ orbifold Wess--Zumino--Witten model at level $2Q_+Q_-$.
This model is the angular sector of a 4D heterotic sigma model where the radial
sector  is a supersymmetric $SL(2,\rline)/U(1)$ coset. The product of these two
models is a bosonised heterotic string theory whose background fields arise as
the extremal limit of the magnetically charged dilaton black hole of
\gibbons\frank. This ends the review.

One of the purposes of this paper is to construct more general 4D heterotic
string backgrounds by mixing the $(r,t)$ and $(\theta,\phi)$ sectors
non--trivially. This is carried out in later sections. The next section will
outline the central and motivating observation about the GPS model:  It is
equivalent to  a gauged WZW model (with fermions) possessing  many interesting
properties.

\newsec{The GPS Monopole theory as a gauged WZW model.}

Now start again by gauging  a right $U(1)$ subgroup of the $SU(2)$ \wzw\ model.
The action of this subgroup (denoted $U(1)_R$) is:
$$
\eqalign{g&\to gh,\jump h=\e{i\alpha\sigma_3/2}\cr \psi&\to\psi+\alpha.}
$$

This gauging procedure is anomalous. However an action for the gauged model may
be written which expresses this (classical) anomaly in such a way that only
depends upon the gauge field\ref\ed{E Witten,  Commun. Math. Phys. {\bf 144},
(1992) 189--212.}:
\eqn\gauged{I(g,A)=I_{WZW_k}+{k\over2\pi}\int d^2z\,\,\Tr [\Azb g^{-1}\d g
]-{k\over4\pi}\int d^2z\,\,\Tr[\Az\Azb].}
Under $\delta g=-gu$ and $\delta A_a=-\partial_au-[A_a,u]$, the non--zero
variation is
\eqn\vary{
\delta I(g,A)={k\over4\pi}\int d^2z\,\,\Tr(uF_{z\bar{z}})\equiv{k\over4\pi}\int
d^2z\,\,\Tr[u(\d\Azb-\db\Az)].
}
For the subgroup defined above, $u=-i\alpha\sigma_3/2$ (and $\delta
A_a=\partial_a\alpha$) and the anomaly is
\eqn\anomaly{\delta I(g,A)={\alpha k\over8\pi}
\int d^2z\,\,F_{z\bar{z}}}
This may be checked directly in the Euler parameterisation of the action. (Note
that  the gauge fields have been written above as $$A_a=iA_a\sigma_3/2.$$ Also
in the Euler parameterisation \euler, $\Tr[\sigma_3 g^{-1}\d
g]=i[\d\psi+cos\theta\d\phi]$.)
This action will be the first building block of the heterotic monopole model.

The rest of this heterotic sigma model is as follows. By asking for (0,1)
supersymmetry there are two right--moving coset  fermions, the supersymmetric
partners of $\theta$ and $\phi$ in the coordinates of the previous sections..
They are minimally coupled to the gauge fields with unit charge. They have an
action
\eqn\fermi{I^R_F={ik\over4\pi}\int d^2z\,\,\,\Tr(\Psi_R{\cal D}_{\bar
z}\Psi_R),}
where $\Psi$ takes its values in the Lie algebra of $SU(2)/U(1)_R$ and
${\cal D}_{\bar z}$ is the gauge covariant derivative. This action is
classically gauge invariant under the above gauge transformations. However,
there is the familiar quantum anomaly at one loop:
\eqn\Fanomr{\delta I^R_F={\alpha\over4\pi}\int d^2z\,\, F_{z\bar{z}}.}

For the left--moving sector   add two left--moving   fermions which will play
the role of the current algebra fermions in the   heterotic string theory. They
will be coupled to the two--dimensional gauge field with strength $Q$  and will
contribute an anomaly:
\eqn\Fanoml{\delta I^L_F={-Q^2\alpha\over4\pi}\int d^2z\,\, F_{z\bar{z}}.}

Notice that all of the anomalies discussed have the same structure. Therefore,
 an anomaly--free gauge invariant theory $$I^{total}=I(g,A,k)+I^R_F+I^L_F$$
may be constructed if the anomalies cancel, that is if
\eqn\cancel{k=2Q^2-2=2Q_+Q_-.}
  This is the same condition as in the model of GPS.

With the above ingredients it is difficult to carry out the remaining
procedures involved in the construction of the coset theory. Although the model
has been shown to be gauge invariant by considering it at one loop, the
Lagrangian constructed out of the fermions and the WZW model is not classically
 gauge invariant, and in this form will not yield the correct coset theory, if
(for example) the gauge fields were integrated out\foot{Typically the quadratic
terms in the gauge fields will not have the correct form.}. One way to proceed
is to bosonise the fermions. By doing this, the one--loop  anomalies become
classical anomalies, and thus $I_{total}$ can be written as a manifestly gauge
invariant theory at tree level.

Now   the fermions bosonise into a $2\pi$ periodic bosonic field $\Phi$ with
the following action:
\eqn\Ii{I_B={1\over4\pi}\int d^2z\,\,\,(\d\Phi-Q_+\Az)(\db\Phi-Q_+\Azb)
-Q_-\Phi F_{z\bar{z}}.
}
The term proportional to $F_{z\bar{z}}$ will yield the anomaly of the fermionic
theories, now made classical by the bosonisation procedure.
The total action  $I_{total}$ has the following $U(1)_R$ gauge  invariance
\eqn\inv{\eqalign{\delta\psi&=\alpha\cr\delta\Phi&=Q_+\alpha\cr\delta
A_a&=\partial_a\alpha.}}

Next fix a gauge and integrate out the gauge fields. There are (at least) two
interesting gauge choices:

\eqn\choice{\eqalign{{\rm (1)} \hskip1cm\Phi&=0;\cr
{\rm (2)}\hskip1cm\psi&=\mp\phi.}}

Condition  (1) will return directly to the $SU(2)/Z_{2Q_+}$ theory of GPS. With
$\Phi$ gauged to zero the earlier identification
$\psi=\xi\mp\phi=X^3/Q_+\mp\phi$ may be made.
The extra identification  $X^3\sim X^3+2\pi$ corresponds to a residual discrete
$Z_{2Q_+}$ subgroup of the original $U(1)_R$ gauge group: The gauge fixing (1)
was incomplete.

In case (2) the following action results (renaming $\Phi$ as $X^3$):
$$\eqalign{I=I_{WZW}+{k\over8\pi}\int d^2z\Biggl\{&{2\over k}\d X^3\db X^3
+\Azb\left(4A^M_\phi\d\phi-{4\over k}\d X^3\right)\cr&+\Az\left(-{4Q\over k}\db
X^3\right)+\Az\Azb\left(1+{2Q^2_+\over k}\right)\Biggr\}}.$$
Finally, integrating out the gauge fields  and recalling \cancel\ leaves
finally the GPS theory:
\eqn\Gps{\eqalign{I_{GPS}={Q_+Q_-\over4\pi}\int d^2z\,\,\,\Biggl\{
(G^{S_2}_{\mu\nu}&+4A^M_\mu A^M_\nu)\d X^\mu \db X^\nu\cr
&+{1\over Q_+^2}(\d X^3-4Q_+A^M_\mu\d X^\mu)\db X^3\Biggr\}
}}  This completes the quantum construction of the `monopole'  sector of the
heterotic sigma model of Giddings, Polchinski and Strominger, as a gauged
\wzw\ model. Note that this  model is actually has $(0,2)$ supersymmetry, as
the $(0,1)$ supersymmetry is enhanced by that fact that $SU(2)/U(1)$ is a
K\"ahler coset\ref\kazama{Y Kazama and H Suzuki, Nucl Phys {\bf B321} (1989)
232\semi Y Kazama and H Suzuki, Phys Lett {\bf B216} (1989) 122.}\ref\edtwo{E
Witten, Nucl Phys {\bf B371} (1992), 191.}.

It is important to note here how to determine the 2D spacetime metric and gauge
fields of the final heterotic string theory. This will be vitally important in
more complicated models.

In more `traditional' gauged WZW model constructions,   there is no necessity
for cancelation of anomalies coming from  the spacetime sector with those of
current algebra fermions. In these cases the spacetime metric would simply be
read off a final action like \Gps\ as
$dS^2=Q^2\{d\theta^2+(\sin^2\theta+4A^M_\phi A^M_\phi)d\phi^2\}$, together with
the $U(1)$ gauge field $QA^M_\phi$ coupling to the left movers. This naive
procedure is incorrect here. The first sign that something is different
here is the fact that the Ricci tensor $R_{\mu\nu}$ and the square of the
electromagnetic field strength $F_{\mu\lambda}F_\nu^{\lambda}$ appear now at
the same order in perturbation theory in the $\beta$--function equations. This
is because perturbation theory in $\alpha^\prime\sim 1/k$ has been replaced by
perturbation theory in $1/Q$ due to the anomaly equation \cancel. Therefore the
effects of the fermions on the metric usually negligible to leading order must
be dealt with here.

Calculating this back--reaction of the fermions on the metric is again a
non--trivial task which must be carried out at one loop. However, bosonisation
is again the key to carrying out the procedure efficiently. This is the more
subtle outcome of the anomaly cancelling technique used to construct the models
in this paper. The bosonisation, presented as a device to enable a classically
gauge invariant Lagrangian to be written, also enables the correction to the
naive metric to be calculated as follows. Looking at the bosonic heterotic
sigma model \Gps, the conserved $U(1)$ current is readily identified as:
$dX^3-2Q_+A^M_\mu dX^\mu$, and factorising the action into the following
form\foot{This form is not unrelated to the Kaluza--Klein form.}:
\eqn\kaluza{\eqalign{I={Q_+Q_-\over4\pi}\int d^2z\,&\Biggl\{
\d\theta\db\theta+\sin^2\theta\d\phi\db\phi\cr
&+{1\over Q_+^2}\left(\d X^3-2Q_+A^M_\mu\d X^\mu\right)\left( \db
X^3-2Q_+A^M_\mu\db X^\mu\right)\cr
&-{2A^M_\phi\over Q_+}\left(\db X^3\d X^\mu-\d X^3\db X^\mu\right)\Biggr\}
}}
which upon refermionisation\foot{Crucially the refermionisation must be done
taking into account the presence of the composite gauge field
$A_a=-2Q_+A^M_\mu\partial_aX^\mu$. See the next section for the  more general
case.}\
is equivalent to:
\eqn\heter{\eqalign{I={Q_+Q_-\over4\pi}\int d^2z\,&\Biggl(
\d\theta\db\theta+\sin^2\theta\d\phi\db\phi\Biggr)\cr
+{1\over2\pi}\int d^2z\, &\Biggl\{
\lambda_R\left( \db-2iA^M_\mu\db X^\mu\right)\lambda_R+\lambda_L\left(
\d-2iQA^M_\mu\d X^\mu\right)\lambda_L\cr
&+{1\over Q_+}\lambda_R\lambda_R\lambda_L\lambda_L
\Biggr\}.
}}
This clearly displays  the  heterotic sigma model with the promised bosonic and
fermionic content. The metric is the round 2--sphere metric with radius $Q^2$
and the gauge field is a $U(1)$ monopole with charge $Q$. This was the starting
point of GPS\GPS.

The rest of this paper deals with   other constructions based on the ideas
presented in this section.  These constructions will yield many interesting
models.

\newsec{Some other 2D theories based on $SL(2,\rline)$.}

The procedures used above to obtain the monopole theory as a gauged WZW model
can be used in wider applications. An obvious one is the closely related
problem of discovering new exact 4D black hole backgrounds which will be
addressed later in this paper. First, another obvious avenue will be explored.
What type of physics results from all of the various gaugings   of
$SL(2,\rline)$ beyond those which yield the 2D black hole\edone\ or Liouville
theory--like models\ref\samsonii{A Alekseev and S Shatashvili, Nucl Phys {\bf B
323}
(1989) 719--733.}? Much more freedom is allowed, typically the symmetry:
\eqn\freedom{g\to\e{\epsilon{\sigma_L/2}}g\,\e{\epsilon{\sigma_R/2}},}
where $g\in SL(2,\rline);
\{\sigma_R,\sigma_L\}\in\{\sigma_3,i\sigma_2,\sigma_1,\sigma_\pm=\sigma_1\pm
i\sigma_2\}$, might be gauged,
producing an anomaly
\eqn\anom{{\cal A}= \epsilon\Tr[\sigma_L^2-\sigma_R^2]{k\over8\pi}\int d^2z
F_{z\bar z}} for the appropriately chosen form of the gauged action as
discussed in the previous section. Supersymmetric right moving fermions and
some `current algebra' left movers with a free charge $Q_L$ (or vice--versa)
can be introduced as before to rescue the model from gauge non--invariance, if
the total anomaly equation is satisfied\foot{The sign of $k$ has been reversed
to achieve a signature of $(-++)$ for the $SL(2,\rline)$ manifold. This is
desirable at least in the case of the black hole theory\edone.}:
\eqn\fixk{-k\Tr[\sigma_L^2-\sigma_R^2]=4(Q_L^2-Q_R^2).} Here $Q_R$ is fixed by
the choice of $\sigma_R$. For the normalisations chosen here   $Q_R$ is
positive and satisfies $Q_R^2=|\Tr\sigma_R^2/2|$.

The 2D models are (after bosonising the fermions to get the correct quadratic
terms):
\eqn\themodel{\eqalign{I= {k\over4\pi}\int d^2z&\,\,\,\Tr(g^{-1}\d g\cdot
g^{-1}\db g)+
ik\Gamma(g)
\cr-{k\over4\pi}\int d^2z&\Biggl\{ \Azb\left(\Tr[\sigma_Rg^{-1}\d g]+{4 \over
k}Q_R\d\Phi\right)+\Az\left(\Tr[\sigma_L\db gg^{-1}]+{4\over
k}Q_L\db\Phi\right)\cr &-{2\over k}
\d\Phi\db\Phi-{1\over2}\Az\Azb\left(\Tr[\sigma_Lg\sigma_Rg^{-1}]+
{Q_L\Tr\sigma_R^2-Q_R\Tr\sigma_L^2\over Q_L-Q_R}\right)\Biggr\}.}} They are
gauge invariant under \freedom\ and
\eqn\gaugesymm{\Phi\to\Phi+(Q_L+Q_R)\epsilon,\jump A_a\to
A_a+\partial_a\epsilon} and have $(0,1)$
supersymmetry\foot{As in the $SU(2)$ case in the previous section, this is
enhanced to $(0,2)$ supersymmetry when the K\"ahler conditions for the coset
are satisfied. This will be true for   models based on gauging all of the
subgroups except those generated by the strictly triangular $\sigma_\pm$.}\
(after refermionisation via $\d\Phi\sim k\lambda^1_R\lambda^2_R,\db\Phi\sim
k\lambda^1_L\lambda^2_L$).

These models may be studied in their own right as 2D models, or used as
building blocks for higher dimensional theories, as will be carried out later
in the paper.
To obtain a leading  order approximation to the sigma--model geometry of these
models, the gauge fields may be integrated out to give:
\eqn\leading{\eqalign{I=I_{WZW}-
{k\over2\pi}&\int  d^2z\,{1\over D}
\left\{\Tr[\sigma_Rg^{-1}\d g]\Tr[\sigma_L\db g g^{-1}]\right\}\cr
+{1\over2\pi}&\int d^2z{1\over D}\Biggl\{\left({D}-{8Q_LQ_R\over
k}\right)\d\Phi\db\Phi\cr-&4\left(Q_R\d\Phi\Tr[\sigma_L\db
gg^{-1}]+Q_L\db\Phi\Tr[\sigma_R g^{-1}\d g]\right)\Biggr\}
,}}
where
\eqn\where{D=\Tr[\sigma_Lg\sigma_Rg^{-1}]+
{Q_L\Tr\sigma_R^2-Q_R\Tr\sigma_L^2\over Q_L-Q_R}  }
Considering the one--loop determinant arising from the integration
measure\ref\buscher{{T H Buscher, Phys Lett {\bf B194} (1987) 59\semi T H
Buscher, Phys Lett {\bf B201} (1988) 466\semi E B Kiritsis, Mod Phys Lett {\bf
A6}, (1991) 2871.}}\ will yield the dilaton:
\eqn\dialatonne{{\hat\Phi}=-{1\over2}\log\left(\Tr[\sigma_Lg\sigma_Rg^{-1}]+
{Q_L\Tr\sigma_R^2-Q_R\Tr\sigma_L^2\over Q_L-Q_R} \right)+{\hat\Phi}_0} where
${\hat\Phi}_0$ is an arbitrary constant.

To proceed, it is natural to parameterise $g$  as follows:
\eqn\paramg{g=\e{\tl{\sigma_L/2}}G\e{\tr{\sigma_R/2}}}
such that the gauge transformations \freedom\ act as shifts of $\tl$ and $\tr$
and $G$ is a one--parameter  $SL(2,\rline)$ subgroup chosen to supply the
(gauge invariant) third independent coordinate on the group manifold. $G$
should be chosen  to be in a different conjugacy class from that of either of
the nieghbouring factors in order to   define an independent third coordinate.
This then corresponds to  the parameterisation of the $SL(2,\rline)$ manifold
by doing combinations of $SO(1,1)$ Lorentz boosts, $SO(2)$ rotations and $E(1)$
translations.
The parameters will have  ranges $[0,\infty)$ or $(-\infty,\infty)$ for the
non--compact subgroups and $[0,4\pi]$ for the rotations.
Choosing $G=\exp(r\sigma_r/2)$ where
$\sigma_r\in\{\sigma_3,i\sigma_2,\sigma_1,\sigma_\pm\}$ the coordinate
invariant expressions reduce to:
\eqn\metric{\eqalign{\Tr[g^{-1}\d g\cdot g^{-1}\db g]
=
{1\over4}\biggl(\d r\db r&\,\Tr\sigma_r^2
+\d\tl\db\tl\,\Tr\sigma_L^2+\d\tr\db\tr\Tr\sigma_R^2\cr
+&(\d\tl\db\tr+\d\tr\db\tl)\Tr[\sigma_LG\sigma_RG^{-1}]\biggr)}}
and
\eqn\veirbiens{\eqalign{\Tr[\sigma_Rg^{-1}\d
g]&={1\over2}(\d\tr\Tr\sigma_R^2+\d\tl\Tr[\sigma_LG\sigma_RG^{-1}]) \cr
\Tr[\sigma_L\db gg^{-1}]&=
{1\over2}(\db\tl\Tr\sigma_L^2+\db\tr\Tr[\sigma_LG\sigma_RG^{-1}]).}}
As $SL(2,\rline)$ is three dimensional the  unique choice (up to scalings) of
the integrand of the 3D Wess--Zumino term is the volume form $\omega_V$
obtained by simply reading off the metric $M$ from \metric\ and forming
$\omega_V=|{\rm det}M|^{1/2}dr\wedge d\tl\wedge d\tr$, where
\eqn\detm{{\rm det}M=\Tr\sigma_r^2\left(\Tr\sigma_L^2\Tr\sigma_R^2
-\Tr[\sigma_LG\sigma_RG^{-1}]\right)}   is clearly a function of $r$ only.
So it is possible to write the volume form as a closed form $\omega_V=d\lambda$
and solve uniquely for the components of $\lambda=\lambda_{RL}(r)d\tr\wedge
d\tl$ in terms of the first $r$--integral of $|{\rm det}M|^{1/2}$. Depending
upon the choices made for   $G$,   the solution will be able to be written
globally or only locally. Either way, the torsion term of the sigma model can
be written as a two--dimensional field theory due to the choices made above.

Given the choice \paramg\ it is   natural to work in terms of the gauge
invariant combination $t=\pm(\tl-\tr)$ by choosing a gauge in which either
$\tr$ or  $\tl$ (respectively) is zero. In this gauge the Wess--Zumino term
disappears, and the  final theory is:
\eqn\modelfinal{\eqalign{I={k\over2\pi}\int d^2z\Biggl\{
{\Tr\sigma_r^2\over8}\,&\d r\db r
-{\Tr\sigma_R^2\over8}\,\d t\db t
{\Tr[\sigma_LG\sigma_RG^{-1}]-\left({Q_L\Tr\sigma_R^2-Q_R\Tr\sigma_L^2\over
Q_L-Q_R}  \right) \over \Tr[\sigma_LG\sigma_RG^{-1}]
+\left({Q_L\Tr\sigma_R^2-Q_R\Tr\sigma_L^2\over Q_L-Q_R}\right)}\cr
&+\left({1\over k}-{8Q_LQ_R\over k^2\left[\Tr[\sigma_LG\sigma_RG^{-1}]
+\left({Q_L\Tr\sigma_R^2-Q_R\Tr\sigma_L^2\over Q_L-Q_R}\right)\right]}
\right)\d\Phi\db\Phi\cr
&-{2Q_R\Tr[\sigma_LG\sigma_RG^{-1}]\over k\left[\Tr[\sigma_LG\sigma_RG^{-1}]
+\left({Q_L\Tr\sigma_R^2-Q_R\Tr\sigma_L^2\over Q_L-Q_R}\right)\right]}\db
t\d\Phi\cr
&-{2Q_L\Tr\sigma_R^2\over k\left[\Tr[\sigma_LG\sigma_RG^{-1}]
+\left({Q_L\Tr\sigma_R^2-Q_R\Tr\sigma_L^2\over Q_L-Q_R}\right)\right]}
\d t\db\Phi
\Biggr\}}} together with
a dilaton given by \dialatonne. Also, when the WZW anomaly does not cancel, the
level $k$ is given by \fixk. Recall also that $Q_R$ is fixed by supersymmetry
and is given below equation \fixk.

When it comes to the determination of the metric of the underlying heterotic
string theory the situation is the same as in the previous section. The
$\beta$--function perturbation expansion parameter is $1/Q_L$.
Generically, the metric is of order $Q_L^2$ and as the gauge field coupling to
the left mover $\db\Phi$ is of order $Q_L$, the curvature, dilaton and gauge
field strength terms are all of the same order. The naive metric read off from
\modelfinal\ has non--negligible corrections due to the back--reaction from the
fermions. Again, the present   bosonised form of the model allows an efficient
determination of the correct spacetime metric, by writing it in a form which
prepares it for  re--fermionisation.
The model is of the form:
\eqn\modelformi{\eqalign{I={k\over2\pi}\int d^2z\Biggl\{G^0_{rr}\d r\db
r+&G^0_{tt}\d t\db t\cr
&+{1\over k}([1+F(r)]\d\Phi\db\Phi+2A_L\d t\db\Phi+2A_R\db t\d\Phi)\Biggr\}}}
which should be written  as:
\eqn\modelformii{\eqalign{I={k\over2\pi}\int d^2z&\Biggl\{G^0_{rr}\d r\db
r+\left(G^0_{tt}-{1\over k}[A_L+A_R]^2\right)\d t\db t+{1\over
k}F(r)\d\Phi\db\Phi\cr
&+{1\over k}\left(\d\Phi+[A_L+A_R]\d t\right)
\left(\db\Phi+[A_L+A_R]\db t\right)\cr
&+[A_L-A_R](\d t\db\Phi-\db t\d\Phi)\Biggr\}}}
By studying the 2D boson--fermion relations in the presence of vector and axial
couplings to a $U(1)$ gauge field $A_\alpha$
 it is easy to read off the $(0,1)$ heterotic sigma model as:
\eqn\modelfinal{\eqalign{I={k \over2\pi}\int d^2z\,&\Biggl(
G_{rr}\d r\db r+G_{tt}\d t\db t\Biggr)\cr
+{1\over2\pi}\int d^2z\, &\Biggl\{
\lambda_R\left( \db-2i\Omega_t\db t\right)\lambda_R+\lambda_L\left(
\d-2iQ_LA_t\d t\right)\lambda_L\cr
&+ {1\over2}F_{rt}\psi_R^r\psi_R^t\lambda_L\lambda_L
\Biggr\},
}}
where $G^0_{rr},G^0_{tt},F(r),A_L$ and $A_R$ are all read off from \modelfinal\
using \modelformi\ and the metric, gauge field, tangent space connection and
four--fermi term are given by:
\eqn\metricetc{\eqalign{
G_{rr}=&G^0_{rr}\cr
G_{tt}=&G^0_{tt}-{1\over k}[A_L+A_R]^2\cr
A_t=&A_L\cr
\Omega_t=&A_R\cr
F_{rt}\psi_R^r\psi_R^t=&F(r)\lambda_R\lambda_R;
}}
The dilaton contribution remains the same as in \dialatonne.

As a quick check, it is useful to recover a familiar case.
  When $\sigma_R=\sigma_L=\sigma_3$ then $G$ should be chosen as
$\exp{r\sigma_1/2}$ and $Q_R=1$. Then the WZW model is anomaly free for
arbitrary $k$ without recourse to the fermions which must now be anomaly--free
on their own. This     amounts to the choice $Q_L^2=1$. Notice that for this
case (and any case where the WZW anomaly is zero) the spacetime metric is
simply the naive metric which would be deduced prior to re--fermionisation, as
the contribution to the $\beta$--function from the gauge field is suppressed by
$1/k$. Choosing $Q_L=-1$ yields the solution:
\eqn\asolutioni{\eqalign{dS^2=&{k\over4}\left(dr^2-\tanh^2\left({r\over2}
\right)dt^2\right),\cr
{\hat\Phi}=&-\log\cosh\left({r\over2}\right)+{\hat\Phi}_0,\cr
A_t=&{2\over\cosh(r)+1},}}
which is the Lorentzian 2D black hole\edone\ with a fixed charge  $U(1)$ gauge
field (assuming that $k$ has been fixed to yield the desired  value of the
central charge). The choice $\sigma_R=\sigma_L=i\sigma_2$ reverses the sign on
the timelike component of the metric displayed above, $t$ is now a compact
coordinate and the cigar shaped  metric of the Euclidean black hole is
recovered.
Those two exact charged solutions were first displayed by Ishibashi, Li and
Steif\ref\ishibashi{N Ishibashi, M Li and A Steif, Phys Rev Lett {\bf 67}
(1991) 3336.}\ as a bosonic solution\foot{The attentive reader will note that
their  solution is in a different gauge arising from their different gauge
fixing choice. Their solution is related to the one here by simply adding a
constant to $A_t$.}. The $U(1)$ boson coupled to the theory there to carry the
background gauge field had a free $U(1)$ charge which gave the black hole an
arbitrary charge. Here the right movers' charge has been chosen in order to
yield $(0,1)$ supersymmetry, and because the WZW anomaly is zero this fixes the
left movers' charge also. This is of course not a necessary requirement here.
However, including supersymmetry will be essential later for constructing 4D
heterotic string solutions.

Now it is interesting to move slightly away from the familiar case.
One way to introduce an arbitrary charge on the 2D black hole is to use a
one--parameter deformation of the charged black hole  theories above which may
be obtained by simply taking $\sigma_R=\delta\sigma_L=\delta\sigma_3.$ The
above choice for $G$ is still appropriate here. This yields the solution (for
large $Q$ and rescaling $\delta t\to t$):
\eqn\asolutionii{\eqalign{d\tau^2=&{k\over4}\left[dr^2-
{\cosh^2(r)- 1\over (\cosh(r)+ \delta)^2 } dt^2\right]\cr
{\hat\Phi}=&-{1\over2}\log\left( \cosh(r)+\delta\right)+{\hat\Phi}_0,\cr
A_t=&   -{2Q \over\cosh(r)+\delta },\cr
k=&2{Q^2\over(\delta^2-1)}.}}
This family of Lorentz signature solutions\foot{These solutions may be embedded
in a larger conformal field theory in order to allow a sensible ($U(1)$ charge
independent) central charge to be chosen.}\ has an arbitrary $U(1)$ charge due
to the inclusion of the free parameter $\delta$. In two dimensions this is
calculated by simply taking the large $r$ limit of the scalar $${\tilde
F}=\e{2({{\hat\Phi}-{\hat\Phi}_0})}\epsilon^{\alpha\beta}F_{\alpha\beta},$$
which yields $$Q_E=Q.$$
The solutions also possess interesting physical behaviour at various radii, for
generic values of $Q_E$: The curvature scalar, given by
$R={G_{tt}^{\prime\prime}/ G_{tt}}-(G_{tt}^\prime)^2/2$  shows that there is a
curvature singularity on the surface given by $\cosh(r)=-\delta$. There is also
a horizon at the radius given by
$\cosh^2r=1.$   This is in direct analogy with the original 2D black hole as
described by Witten\edone. There $G_{tt}=\tanh^2{r/2}=(\cosh r-1)/(\cosh r+1)$
and so there is a horizon at $\cosh r=1$ and a singularity at the (analytically
continued) position $\cosh r =-1$. In the present solution, to leading order in
$Q$, the position of the singularity is a function of $\delta.$

Notice that when $\delta=-1$, $k$ is free again, the expression for the metric
\asolutionii\ is invalid (due to the use of \fixk\ in its derivation) and the
familiar trumpet shaped solution dual\foot{Here duality refers to the  metric
of the bosonic theory alone\edone\verlinde. For all of the theories of this
paper the full conformal field theory including the fermions will be dual to a
much richer class of theories than those obtained by   WZW models with no
additional fermions.  It would be interesting to study them.}\ to \asolutioni\
is obtained. Of course, the same construction could have been carried out with
the Euclidean theory $\sigma_R=\delta\sigma_L=i\delta\sigma_2.$

Now it is interesting to turn to the unusual case where the left and right
generators of the $U(1)$ action on the group manifold are in different
conjugacy classes.
In close analogy with the previous case taking
$\sigma_L=\sigma_3,\sigma_R=i\delta\sigma_2,\sigma_r=\sigma_1$ yields  the
solution:
\eqn\asolutioniii{\eqalign{dS^2=&{k\over4}\left[dr^2+
{\sinh^2(r)+ 1\over(\sinh(r)+\delta)^2}dt^2\right],\cr
{\hat\Phi}=&-{1\over2}\log(\sinh(r)+\delta)+{\hat\Phi}_0,\cr
A_t=&{2Q\over\sinh(r)+\delta}\cr
k=&-2{Q^2\over(\delta^2+1)}.}}  The coordinate $t$ in now compact with period
$4\pi$.  This solution has a number of interesting properties.
The most obvious one is the singularity at the   radius given by
$ \sinh r =-\delta$ and a horizon at $\sinh r = 1$. Also note that as
$r\to\infty$ the metric approaches that of a cylinder. Analytically continuing
to  non--compact geometry, this means that the solution is asymptotically flat.

Notice that the cases where both $\sigma_L$ and $\sigma_R$ are either of
$\sigma_\pm$ there is no supersymmetry, as the right--moving fermions decouple.
The WZW anomaly is already zero, so all of the fermions may be neglected. As is
familiar now\verlinde\samsonii\ref\ali{M Alimohammadi, F Ardalan and H Arfaei,
{\sl `Nilpotent gauging of $SL(2,\rline)$ WZNW models and Liouville Field'},
BONN--HE--93--12, SUTDP--93/72/3, IPM--93--007, hep-th/9304024.}, the resulting
theory is a 1D model of the  Liouville form, with cosmological coupling $\mu$,
which in the straightforward case here is zero, but may be made non--zero by
modifying the constraints on the $\sigma_\pm$ generators.

By putting $\sigma_L=\sigma_\pm$ but leaving $\sigma_R$ as one of the other
generators, the right moving supersymmetry is retained and more non--trivial 2D
geometries can be obtained. For  example $\sigma_L=\sigma_+,
\sigma_R=\delta\sigma_3$ and $\sigma_r=\sigma_1$ results in
\eqn\asolutioniv{\eqalign{dS^2=&{k\over4}\left[dr^2-{\sinh^2 (2r)+1 \over
(\sinh(2r)+2\delta)^2} dt^2\right],\cr
{\hat\Phi}=&-{1\over2}\log(\sinh(2r)+2\delta)+{\hat\Phi}_0,\cr
A_t=&-{2Q\over\sinh(2r)+2\delta}\cr
k=&{2Q^2\over(\delta^2-1)}.}} Here $-\infty<t<\infty$.
The solution is asymptotically  flat at   $r=\infty$  and as before for generic
$Q$ there is a horizon and a singularity. This case is similar to the previous
one and may be thought of as the Lorentzian version of it, after a trivial
rescaling of $r$ and $\delta$.

Again, these models can in principle be arranged as part of a larger conformal
field theory where the central charge can be set so as to fix the space time
dimension and leave $Q_E$ free.

\newsec{Charged two--dimensional black holes in string theory.}

The  previous section described a family of conformal field theories which may
be  interpreted as low dimensional charged black hole solutions of string
theory, to one loop in the $\alpha^\prime$ expansion. This means that they are
solutions to the $\beta$--function equations:
\eqn\betafunction{\eqalign{
R_{\mu\nu}+2\nabla_\mu\nabla_\nu{\hat \Phi}-2F_{\mu\lambda}F_\nu^\lambda=0\cr
\nabla^\mu(\e{-2\hat\Phi}F_{\mu\nu})=0\cr
4\nabla^2{\hat\Phi}-4(\nabla{\hat\Phi})^2+\Lambda+R-F^2=0.
}}
The  black hole solutions to these equations (for $F_{\mu\nu}=0$) have been
studied directly by Mandal, Sengupta and Wadia\ref\mandal{G Mandal, A M
Sengupta and S R Wadia, Mod Phys Lett, {\bf A6} (1991) 1685}. Witten's exact
conformal field theory  derived from the $SL(2,\rline)/U(1)$ coset coincided
with this solution at one loop, as indeed it should. McGuigan, Nappi and
Yost\ref\chiara{M D McGuigan, C R Nappi and S A Yost, Nucl Phys {\bf B375}
(1992) 421. hep-th/9111038.} generalised the solution of ref.\mandal\
 to the case where gauge fields are present,
introduced either by coupling to an open string sector
or by heterotic compactification of closed strings. This latter case should be
of concern here in order to interpret the results of the previous section.

In general, the solution takes the following form:
\eqn\chargedbh{\eqalign{dS^2\sim&\,\,
\,(1-2m\e{-Qr}+q^2\e{-2Qr})^{-1}dr^2-(1-2m\e{-Qr}
+q^2\e{-2Qr})dt^2\cr
A_{t}\sim&\,\,\,-\e{-Qr},}}
with linear dilaton
\eqn\linear{{\hat \Phi}-{\hat \Phi}_0=-{Qr\over2}.}

Here, $Q$ is a constant, $q$ sets the charge of the black hole and $m$ is
related to the
physical  mass. The curvature singularity is located at $r=-\infty$, the
solution is asymptotically flat for $r\to\infty$ and there are horizons at the
 zeros of $G_{tt}$. For $q=0$  (the uncharged case) the coordinate
transformation \eqn\transi{\e{Qr}=m(\cosh\sigma+1)} yields
(after absorbing $Q$ into
$\sigma$)
\eqn\wittens{\eqalign{dS^2\sim&\,\,\, d\sigma^2-\tanh^2{\sigma\over2}dt^2\cr
{\hat\Phi}-{\hat\Phi}_0=&-{1\over2}\log(\cosh^2\sigma)},}
 which is the familiar form of the 2D black hole.

Similarly, the transformation
\eqn\transii{\e{Qr}={m\over\delta}(\cosh\sigma+\delta)} produces (for
$q^2=m^2(\delta^2-1)/\delta^2$)
\eqn\mine{\eqalign{dS^2\sim&\,\,\,
d\sigma^2-{\cosh^2\sigma-1\over(\cosh\sigma+\delta)^2}dt^2\cr
{\hat\Phi}-{\hat\Phi}_0=&-{1\over2}\log(\cosh\sigma+\delta)\cr
A_t\sim&\,\,-{1\over(\cosh\sigma+\delta)}
},}
which is the solution \asolutionii\ of the previous section. Furthermore, the
solutions \asolutioniii\ and \asolutioniv\ may be obtained by similar
transformations, replacing $\cosh\sigma$ by $\sinh\sigma$, etc. This explains
why those solutions  all appear to be simply analytic continuations of each
other.

The conclusion here is that the interpretation of the conformal field theories
explicitly examined in the previous section is indeed the 2D heterotic string
 theory in a black hole background with $U(1)$ gauge field.
In the leading order
approximation, the different solutions found are different coordinate
transformations of the same basic solution presented  in ref.\chiara.
The mass and charge of the 2D black hole, which are independent in general,
 have been set
 proportional to one another\foot{Generically there is only one free parameter
 left in the theories of the previous section. }. Note however that this is
not analogous to the familiar 4D  extremality condition as the singularity and
horizon are
still distinct. The details of the spacetime structure of the charged 2D black
 holes was analysed in ref.\chiara.

As mentioned in the introduction, extremal 4D black holes factorise into  a 2D
radial theory and a 2D angular theory.
Section 3 discussed how to realise a magnetic monopole angular theory
background of heterotic string theory as a gauged WZW model with fermions.
Section 4 described radial conformal field theories of electrically charged
linear dilaton black holes  by the same construction. Arbitrary products of
these two types of theories can be made to realise extremal 4D black holes with
$U(1)\times U(1)$ gauge group where one $U(1)$ factor carries the electric
charge and the other the magnetic.

It would be interesting however to construct less trivial combinations of the
radial and angular degrees of freedom, using the same techniques. This would
yield exact conformal field theories of more interesting 4D backgrounds. This
is the subject of the rest of this paper. The result will be  a $U(1)\times
U(1)$ dyon, where both $U(1)$ factors contain electric and magnetic components.
The metric will not be a product metric.

%the rest.

\newsec{A gauged $ SL(2,\rline)\times SU(2) $ \wzw\ model.}
The construction  starts by gauging certain $U(1)$  subgroups of an
$SL(2,\rline)\times SU(2)$ \wzw\ model:
\eqn\begin{\eqalign{I_{WZW}=
{k_1\over4\pi}\int_\Sigma d^2z\,\,\,&\Tr(g_1^{-1}\d g_1\cdot g_1^{-1}\db g_1)
-{k_2\over4\pi}\int_\Sigma d^2z\,\,\,\Tr(g_2^{-1}\d g_2\cdot g_2^{-1}\db
g_2)\cr &+ik_1\Gamma(g_1)-ik_2\Gamma(g_2)
}}
where
\eqn\gam{\Gamma(g)=\int_B g^*\omega.
}
 Here $g_1\in SL(2,\rline)$ and $g_2\in SU(2)$.
This model is a theory of maps from $\Sigma$ to $G=SL(2,\rline)\times SU(2)$.
$\Sigma$ is ($1+1$ dimensional) spacetime, the boundary of an auxiliary space
$B$.
Here $\omega$ is a $G_L\times G_R$ invariant 3--form on $G$:
\eqn\omeg{\omega={1\over12\pi}\Tr(g^{-1}dg\wedge g^{-1}dg\wedge g^{-1}dg)}

The first subgroup of interest is
\eqn\combined{U(1)_A\times U(1)_B:\left\{\jump\eqalign{
g_1\to&\e{\epsilon_A\sigma_3/2}g_1
\e{(\delta\epsilon_A+\lambda\epsilon_B)\sigma_3/2}\cr
g_2\to&g_2\e{i\epsilon_B\sigma_3/2}   }\right.} (Notice that the limit
$\lambda\to0,\delta\to1$ will yield the gauging for the monopole theory of the
 section 3.)

As in the previous construction gauging this subgroup will result in anomalies.
It is not possible to construct a gauge invariant extension of the Wess--Zumino
term $\Gamma(g)$. It is possible however to construct an extension
$\Gamma(g,A^A,A^B)$ for which the terms violating gauge invariance do not
depend upon the group element\ed:
\eqn\gaugeext{\eqalign{\Gamma(g,A^A,A^B)=\Gamma(g)-{1\over4\pi}\int_\Sigma
d^2z\,\,\, A^a\wedge\Tr(t_{a,L}dgg^{-1}+t_{a,R}g^{-1}dg)\cr
-{1\over8\pi}\int_\Sigma d^2z\,\,\, A^a\wedge
A^b\Tr(t_{a,R}g^{-1}t_{b,L}g-t_{b,R}g^{-1}t_{a,L}g)
,}}
where the indices $a,b$ can take the values $A,B$. Also $A^A$ and $A^B$ are
gauge  fields for $U(1)_A$ and $U(1)_B$ respectively, and $t_{a,L(R)}$ are the
generators for the left(right) action of the groups.
Under gauge transformations this action will have anomalies
$$
{\cal A}_{ab}={1\over4\pi}\Tr(t_{a,L}t_{b,L}-t_{a,R}t_{b,R})\int_\Sigma
d^2z\,\,\,F^b_{z\bar z}
$$
where $F^b_{z\bar z}=\d\Azb^b-\db\Az^b$ and there is no sum intended on the
index $b$ in the formula immediately above.
The metric part of the \wzw\ model is coupled invariantly to gauge fields by
simply replacing the derivative with its gauge--covariant extension:
$$
\partial_\mu g\to{\cal D}_\mu g=\partial_\mu g+A^a_\mu(t_{a,L}g-gt_{a,R}).
$$
Inserting the explicit expressions for the generators:
$$
t^{(1)}_{A,R}=\delta{\sigma_3\over2};\,t^{(1)}_{A,L}={\sigma_3\over2};
\,t^{(1)}_{B,R}=
\lambda{\sigma_3\over2};\,t^{(2)}_{B,R}=i{\sigma_3\over2},
$$
yields the following total action for the gauged \wzw\ model:
\eqn\totalaction{\eqalign{I=I_{WZW}
+&{k_1\over8\pi}
\int
d^2z\,\,\Biggl\{-2\left(\delta\Azb^A+\lambda\Azb^B\right)\Tr[\sigma_3g_1^{-1}\d
g_1]-2\Az^A\Tr[\sigma_3\db g_1g_1^{-1}]\cr
&\jump\jump+\Az^A\Azb^A\left(1+\delta^2+\delta
\Tr[\sigma_3g_1\sigma_3g_1^{-1}]\right)+
\lambda^2\Az^B\Azb^B\cr
&\jump\jump+\delta\lambda\Az^A\Azb^B+
\Az^B\Azb^A
\left(\delta\lambda+\lambda\Tr[\sigma_3g_1\sigma_3g_1^{-1}]\right)\Biggr\}\cr
+&{k_2\over8\pi}\int d^2z\,\, \Biggl\{2i\Azb^B\Tr[\sigma_3g_2^{-1}\d
g_2]+\Az^B\Azb^B\Biggr\}
}}
with anomalies
\eqn\anomalys{\eqalign{{\cal A}_{AA}=&-{k_1(1-\delta^2)\epsilon_A\over8\pi}\int
d^2z\,F^{A}_{z\bar{z}}\cr
{\cal A}_{BB}=&(k_2+k_1\lambda^2){\epsilon_B\over8\pi}\int
d^2z\,F^{B}_{z\bar{z}}\cr
{\cal A}_{AB}=&k_1\delta\lambda{\epsilon_A\over8\pi}\int
d^2z\,F^{B}_{z\bar{z}}\cr
{\cal A}_{BA}=&k_1\delta\lambda{\epsilon_B\over8\pi}\int
d^2z\,F^{A}_{z\bar{z}}.}}

Turning to the fermionic content,  consider the   four $(0,1)$ right-moving
supersymmetric coset fermions. They are minimally coupled to the above gauge
theory:
\eqn\Rminimally{I^F_R={i\over4\pi}\int d^2z\,\,\Tr(\Psi_R{\cal D}_{\bar
z}\Psi_R).}
Here $\Psi_R$ takes values in the orthogonal complement of the Lie algebra of
$U(1)_A\times U(1)_B$ and the covariant derivative is
\eqn\cov{{\cal D}_{\bar z}\Psi_R=\db\Psi_R-\sum_aA^a_{\bar z}[t_{a,R},\Psi_R]}
where the $t_{a,R}$ are the generators of the subgroup acting on the  right.
As the generators of the $U(1)$ subgroups of $SL(2,\rline)$ and $SU(2)$ are
represented as diagonal antihermitian $2\times2$ matrices, the fermions are of
the form:
\eqn\ferms{\eqalign{&\Psi_{R,1}= \pmatrix{0&\lambda^1_R\cr\lambda^2_R&0},\cr
&\Psi_{R,2}=\pmatrix{0&\lambda^3_R\cr\lambda^4_R&0}}}  with `1' and `2'
denoting
$SL(2,\rline)$ and $SU(2)$ respectively. The $\Tr$ in \Rminimally\ decomposes
as $-k_1\Tr_1+k_2\Tr_2$. The resulting action is therefore:
\eqn\resulting{\eqalign{I^F_R=-&{ik_1\over4\pi}\int d^2z\,\,
\left\{2\lambda^1_R\db\lambda^2_R+\lambda^1_R[\delta\Azb^A
+\lambda\Azb^B,\lambda^2_R]  \right\}\cr
+&{ik_2\over4\pi}\int
d^2z\,\,\left\{2\lambda^3_R\db\lambda^4_R+\lambda^3_R[\Azb^B,\lambda^4_R]
\right\}.}}

The model now has invariance under the $(0,1)$ supersymmetry
\eqn\susan{\eqalign{\delta g_1&=i\epsilon g_1\Psi_{R,2};\cr
\delta g_2&=i\epsilon g_2\Psi_{R,2};\cr
\delta\Psi_{R,1}&=\epsilon(g_1^{-1}\d g_1+{\Az^A\over2}
g_1^{-1}\sigma_3g_1);\cr
\delta\Psi_{R,2}&=\epsilon(g_2^{-1}\d g_2)\cr
\delta A_i^a&=0,}} (modulo the equations of motion
$\d\Psi_{R,1}=0,\d\Psi_{R,2}-{\Az^A}[\sigma_3,\Psi_{R,2}]/2$) which may be
verified by direct calculation. In general this symmetry is enhanced to $(0,2)$
supersymmetry in a general $G/H$ coset model under certain circumstances\edtwo:
The complexification of the  space Lie$(G/H)$ denoted $T_C$ can be decomposed
into the form $T_C=T\oplus \bar T$. Here $T$ and $\bar T$ are represented in
this case by strictly upper   and strictly lower triangular $2\times2$ matrices
respectively. They therefore automatically satisfy the required conditions
\eqn\cond{[T,T]\in T,\jump [{\bar T},{\bar T}]\in {\bar T}} and
\eqn\condd{\Tr(t_1,t_2)=0\jump{\rm for}\,\,\, t_1,t_2\in T\,\, {\rm
or}\,\,{\bar T}.} These are precisely the algebraic conditions for a coset
space to be K\"ahler, which is the requirement for enhancing $N=1$ to $N=2$
supersymmetry as pointed out by Kazama and Suzuki\kazama.  The fermions
$\Psi_R$, which take their values in $T_C$, have been decomposed in this case
as given by equation \ferms.
By inspection of the action \resulting\ it can be seen that it is possible to
assign a  classical\foot{To identify the full $U(1)$ R--symmetry of these
cosets  the classical and quantum   anomalies coming from the bosonic and
fermionic sectors respectively have to be treated correctly. See for example
ref\ref\mans{M Henningson, Nucl Phys {\bf B413} (1994) 73. hep-th/9307040.}.}\
R--symmetry  to the system under which $g_1,g_2$ and the gauge fields have
charge zero,  fields valued in $T$ have charge 1 and fields valued in $\bar T$
have charge $-1$. This symmetry does not commute with the transformations
\susan\ and therefore   a second supersymmetry action may be extracted. The
explicit $(0,2)$ supersymmetry transformations for the fields in this model may
be found by using the decomposition \ferms\    (together with the
parameterisations of $g_1$ and $g_2$ and to be discussed later)  in the
transformations \susan.

Next add four left--moving fermions
$\Psi_L=(\lambda^1_L,\lambda^2_L,\lambda^3_L,\lambda^4_L)$. These carry a
global $SO(4)_L$ symmetry, the maximal torus of which will be identified with
the gauged $U(1)_A\times U(1)_B$. The generators of this gauge symmetry will be
chosen as:
\eqn\generators{{\hat Q}_A=\pmatrix{0&Q_A&0&0\cr -Q_A&0&0&0\cr0&0&0&P_A\cr
0&0&-P_A&0},{\hat Q}_B=
\pmatrix{0&Q_B&0&0\cr -Q_B&0&0&0\cr0&0&0&P_B\cr 0&0&-P_B&0}}
giving  the action:
\eqn\fermileft{\eqalign{I^F_L=-{ik_1\over4\pi}\int d^2z\,\,
\left\{\lambda^1_L[\d+Q_A\Az^A+Q_B\Az^B]\lambda^2_L
\right\}\cr
+{ik_2\over4\pi}\int d^2z\,\,
\left\{\lambda^3_L[\d+P_A\Az^A+P_B\Az^B]\lambda^4_L
\right\}.
}} (Here the couplings $k_1$ and $k_2$ have explicitly been chosen.) Note that
to connect smoothly to the pure monopole theory, the charges $P_A$ and $Q_B$
should be sent to zero as $\delta\to1$ and $\lambda\to0$. (This is necessary to
ensure that the remaining mixed anomaly from the left moving fermions cancel in
this limit. $Q_A$ remains non--zero to cancel the anomaly from the
$SL(2,\rline)$ right--moving fermions.)

At one--loop the fermions  will produce   anomalies of the same form as above
when coupled to gauge fields. They are:
\eqn\fanomalys{\eqalign{{\cal A}^F_{AA}
=&-2(Q_A^2+P_A^2-\delta^2) {\epsilon_A\over8\pi}\int d^2z\,F^{A}_{z\bar{z}}\cr
{\cal A}^F_{BB}=&-2(Q_B^2+P_B^2-(1+\lambda^2 )){\epsilon_B\over8\pi}\int
d^2z\,F^{B}_{z\bar{z}}\cr
{\cal A}^F_{AB}=&-2(Q_AQ_B+P_AP_B-\lambda\delta){\epsilon_A\over8\pi}\int
d^2z\,F^{B}_{z\bar{z}}\cr
{\cal A}^F_{BA}=& -2(Q_AQ_B+P_AP_B-\lambda\delta){\epsilon_B\over8\pi}\int
d^2z\,F^{A}_{z\bar{z}}.}}

Adding these anomalous fermionic actions to the anomalously gauged \wzw\ model
will yield a gauge invariant theory if:
\eqn\cancel{\eqalign{-k_1(1-\delta^2)&= 2(Q_A^2+P_A^2-\delta^2);\cr
k_2+k_1\lambda^2&= 2(Q_B^2+P_B^2-(1+\lambda^2));\cr
k_1\delta\lambda&=2(Q_AQ_B+P_AP_B-\lambda\delta) .
}}
Also, as this is a four--dimensional heterotic string solution, the resulting
central charge of this conformal field theory\foot{The rest of the central
charge needed to construct a consistent string theory will be supplied by a
$c=9$ internal conformal field theory in the usual way. This will not be of
concern here.}\ should be $c=6$. This means that
\eqn\central{c={3k_1\over k_1-2}+{3k_2\over k_2+2}=6,} (the $-2$ from gauging
is cancelled by the $+2$ from the four fermions) which gives
$k_1=k_2+4$.

All these  fermions are equivalent to two periodic bosons $\Phi_{1}$ and
$\Phi_{2}$ which   display the anomalies \fanomalys\ classically:
\eqn\thebosons{\eqalign{
I_B={1\over4\pi}\int d^2z\,\,\,
\Biggl\{&(\d\Phi_2-P_A\Az^A-(P_B+1)A^B_z)^2 \cr
+&(\d\Phi_1-(Q_B+\lambda)A^B_z-(Q_A+\delta)A^A_z)^2\cr
-&\Phi_1\Biggl[(Q_B-\lambda)F^B_{z\bar z}+(Q_A-\delta)F^A_{z\bar z}\Biggr]\cr
-&\Phi_2 \Biggl[(P_B-1)F^B_{z\bar z}+ P_AF^A_{z\bar z}\Biggr]\cr
+&\Biggl[\Azb^A\Az^B-\Az^A\Azb^B\Biggr]\Biggl[\delta Q_B-Q_A\lambda-P_A
\Biggr]\Biggr\}.}}
The  $U(1)_A\times U(1)_B$ action on the bosons is:
\eqn\acta{\matrix{\delta\Phi_1=(Q_A+\delta)\epsilon_A+(Q_B+\lambda)
\epsilon_B&\delta\Phi_2=P_A\epsilon_A+(P_B+1)\epsilon_B;
\cr\cr\delta A^A_a=\partial_a\epsilon_A&\delta A^B_a=\partial_a\epsilon_B,}}
under which it may be verified that the action \thebosons\ yields precisely the
anomalies displayed in \fanomalys.
\vfill\eject

\newsec{The complete theory.}

The complete theory is  now (after using the anomalies to simplify the
quadratic terms in the gauge fields):
\eqn\bigaction{\eqalign{I_{total}=I_{WZW}&+{1\over4\pi}\int d^2z\,\,
\Biggl\{\d\Phi_1\db\Phi_1+\d\Phi_2\db\Phi_2
\Biggr\}\cr
-{k_1\over8\pi}
\int d^2z\,\,\Biggl\{
&\Az^A\left(2 \Tr[\sigma_3\db g_1g_1^{-1}]+{4\over k_1}
(Q_A\db\Phi_1+P_A\db\Phi_2)\right)\cr+
&\Azb^A\left(2\delta\Tr[\sigma_3g_1^{-1}\d g_1]+{4\over k_1}\delta \d\Phi_1
\right)\cr+
&\Az^B\left( {4\over k_1}(Q_B\db\Phi_1+P_B\db\Phi_2) \right)\cr+
&\Azb^B\Biggl(2\lambda\Tr[\sigma_3g_1^{-1}\d g_1]-2i{k_2\over
k_1}\Tr[\sigma_3g_2^{-1}\d g_2]\cr&\jump+{4\over k_1}
(\d\Phi_2+\lambda\d\Phi_1) \Biggr)\cr-
&\Az^A\Azb^B\left(\lambda\Tr[\sigma_3g_1\sigma_3g_1^{-1}]
+ {4\over k_1}
[P_A(P_B+1)+Q_A(Q_B+\lambda)] \right)\cr-
&\Az^B\Azb^A\left( {4\over k_1} [P_AP_B+Q_B(Q_A+\delta)]  \right)\cr
-&\Az^A\Azb^A\left(2+\delta\Tr[\sigma_3g_1\sigma_3g_1^{-1}]+{4\over k_1}
[P_A^2+Q_A(Q_A+\delta)] \right)\cr-
&\Az^B\Azb^B\left(  {4\over k_1} [P_B(P_B+1)+Q_B(Q_B+\lambda)]  \right)
\Biggr\}
}}
Due to the anomaly cancelation \cancel\ it is gauge invariant and hence
conformally invariant, as can be be verified explicitly.
\def\TRA{\Tr_a}\def\TRB{\Tr_b}\def\TRC{\Tr_c}\def\TRD{\Tr_d}

As the gauge fields are non--dynamical and appear quadratically in the action
they may be integrated out with the following result:
\eqn\result{\eqalign{I=I_{WZW}&+\cr
{k_1\over8\pi}\int d^2z {1\over D}&
\Biggl\{{2D\over k_1}[\db\Phi_2\d\Phi_2+\db\Phi_1\d\Phi_1]\cr
-4\TRA\Biggl[(\lambda F_B&-\delta G_B)(\TRB+{2\over
k_1}\d\Phi_1)+F_B(\TRC+{2\over k_1}\d\Phi_2)\Biggr]\cr
+(Q_A\db\Phi_1&+P_A\db\Phi_2)\Biggl[8(\lambda F_B-\delta G_B)(\TRB+{2\over
k_1}\d\Phi_1)+8F_B(\TRC+{2\over k_1}\d\Phi_2)\Biggr]\cr
-(Q_B\db\Phi_1&+P_B\db\Phi_2)\Biggl[8(\lambda F_A-\delta
G_A+\lambda{k_1\over2})(\TRB+{2\over k_1}\d\Phi_1)\cr
&\hskip5cm+(8F_A-2\delta\TRD+4k_1)(\TRC+{2\over k_1}\d\Phi_2)\Biggr]
 \Biggr\},
}}
where
\eqn\denomin{D\equiv\TRD( \lambda F_B-\delta G_B)-2G_B+{4(F_BG_A-G_BF_A)\over
k_1}}
with
$$
\eqalign{F_A=&P_A^2+Q_A(Q_A+\delta);\cr
F_B=&P_AP_B+Q_B(Q_A+\delta);\cr
G_A=&P_A(P_B+1)+Q_A(Q_B+\lambda);\cr
G_B=&P_B(P_B+1)+Q_B(Q_B+\lambda);\cr
\TRA=&{\Tr[\sigma_3\db g_1g_1^{-1}]},\,\,
\TRB={\Tr[\sigma_3g^{-1}_1\d g_1]},\cr
\TRC=&-i{k_2\over k_1}{\Tr[\sigma_3g^{-1}_2\d g_2]},\,\,
\TRD={\Tr[\sigma_3g_1\sigma_3g^{-1}_1]}.}
$$

Due to the unusual construction of this model, using bosonisation to arrive at
the desired quadratic terms for the gauge fields to achieve gauge invariance,
it is worth checking that gauge invariance is present after the integration
process. Using  the gauge transformations \combined\ and  \acta, together with
repeated use of the anomaly equations \cancel,  gauge invariance is indeed
verified.

Naively there is no reason to expect that the integration procedure is correct
beyond the leading order in $k_1$, so the terms appearing in \result\ are the
leading   order terms in a large $k_1$ expansion.  Also   $k_1/k_2\to1$ in this
limit, following from equation \central. In what follows $k_1$ and $k_2$ will
be denoted $k$ and in this limit the anomaly equations  simplify:
\eqn\simpleranomaly{k=2{Q_A^2+P_A^2\over\delta^2-1}=2{Q_B^2+P_B^2
\over\lambda^2+1}=2{Q_AQ_B+P_AP_B\over\delta\lambda}.} It is also extremely
useful   that
\eqn\moresimplifications{\eqalign{F_A&={k\over2}(\delta^2-1),\cr
F_B&=G_A={k\over2}\lambda\delta,\cr
G_B&={k\over2}(\lambda^2+1).}}
It is comforting to note that the potentially clumsy expressions resulting from
\result\ and \denomin\ simplify enormously in this large radius limit using
these expressions, as will be shown in the next section once the model has been
endowed with some coordinates.

\newsec{The 4D solutions}

The next task is to choose a parameterisation for $g_1$ and $g_2$ and
appropriate gauge conditions.
 The choice of parameterisation for    $SL(2,\rline)$ shall be the Euler
angles:
\eqn\euleri{g_1=\e{\tl\sigma_3/2}\e   {\sigma\sigma_1/2}\e{\tr\sigma_3/2}
=\pmatrix{
\e{{t_+\over2} }\cosh   {\sigma\over2}&
\e{{t_-\over2} }\sinh   {\sigma\over2}\cr
\e{-{t_-\over2} }\sinh   {\sigma\over2}&
\e{-{t_+\over2} }\cosh   {\sigma\over2}\cr
},}
where the $\sigma_i$ are the Pauli matrices and
\eqn\rangesi{t_\pm\equiv\tl\pm\tr,\jump
0\leq\sigma\leq\infty,\jump-\infty\leq\tl\leq\infty,
\jump-\infty\leq\tr\leq\infty,}    and Euler angles
$(\phi,\theta,\psi)$ for $SU(2)$ are chosen as in \euler\ and
\ranges. For these choices    the traces are:
\eqn\traces{\eqalign{&{\Tr[\sigma_3\db
g_1g_1^{-1}]}=\db\tl+\db\tr\cosh\sigma,\cr
&{\Tr[\sigma_3g^{-1}_1\d g_1]}=\d\tr+\d\tl\cosh\sigma,\cr
&{\Tr[\sigma_3g_1\sigma_3g^{-1}_1]}=2\cosh\sigma,\cr
&\Tr[\sigma_3g^{-1}_2\d g_2]=i(\d\psi+\d\phi\cos\theta).}}
 The gauge transformations in these coordinates are:
\eqn\transform{\eqalign{&\tl\to\tl+\epsilon_A;\cr
&\tr\to\tr+\delta\epsilon_A+\lambda\epsilon_B;\cr
&\psi\to\psi+\epsilon_B;\cr
&\Phi_1\to\Phi_1+(Q_A+\delta)\epsilon_A+(Q_B+\lambda)\epsilon_B;\cr
&\Phi_2\to\Phi_2=P_A\epsilon_A+(P_B+1)\epsilon_B.}}

There are three   gauge choices of interest  which may be chosen for this
model:
\eqn\morechoices{\eqalign{{\rm (1)}\jump\Phi_1&=\Phi_2=0;\cr
{\rm (2)}\jump\tl&=0,\,\,\psi=\mp\phi;\cr
{\rm (3)}\jump\tl&=\tr=0.}}

Choices (1) and (2) are  equivalent (up to a simple change of variables) for
all values of the parameters, in analogy with the two gauge choices discussed
for the monopole theory in section 3. With gauge (1),  $\tr$ and $\psi$  can be
identified as the bosonised world sheet fermions, after a discrete
identification to obtain   $2\pi$ periodicity. This modding is the analogue of
that in the GPS monopole theory.

The gauges (3) and (2) correspond respectively to whether the  gauge conditions
are applied entirely to the $SL(2,\rline)$ coordinates or shared between the
$SL(2,\rline)$ and $SU(2)$ coordinates. These gauges are also equivalent up to
simple coordinate transformations away from $\lambda=0$, where gauge (3) cannot
be implemented. However when $\lambda=0$ the angular and radial sectors of the
theory decouple; the angular sector is simply the monopole theory of section 3
while the radial sector is the 2D charged black hole model of sections 4 and 5.

Choosing gauge (2) (as it is the most transparent and generic) and examining
the gauge transformations \transform\ the remaining fields and coordinates are
defined in terms of the gauge invariant combinations of the original fields:
\eqn\coords{\eqalign{
&\sigma,\theta,\phi,\cr
&t^\prime=\tr-\delta\tl-\lambda(\psi\pm\phi),\cr
&\Phi_1^\prime=\Phi_1-(Q_A+\delta)\tl-(Q_B+\lambda)(\psi\pm\phi),\cr
&\Phi_2^\prime=\Phi_2-P_A\tl-(P_B+1)(\psi\pm\phi).
}}
Having chosen the $U(1)_B$ gauge, some care must be exercised to trace the
global $SU(2)_L$ invariance in the now gauge--fixed coordinates.  An $SU(2)_L$
rotation preserves the two--sphere defined by $(\theta,\phi)$ and the $U(1)$
monopole field undergoes a gauge transformation:
\eqn\mono{A^M={(\pm1-\cos\theta)\over2}d\phi\to A^M+d\Lambda.}
$SU(2)_L$ invariance is preserved by the fact that the forms
$dt^\prime,d\Phi_1^\prime,d\Phi_2^\prime$ all shift by amounts proportional to
$d\Lambda$. This structure is guaranteed to preserve $SU(2)_L$ invariance as it
is inherited directly from the  world sheet $U(1)_B$ gauge theory.

Dropping the primes the model may be written as follows:
\eqn\aform{\eqalign{I={k_1\over8\pi}\int d^2z& \Biggl\{
G_{\sigma\sigma}\d\sigma\db\sigma+G^0_{tt}\d t\db t+G^0_{t\phi}(\d t\db\phi+\db
t\d\phi)\cr
&+G_{\theta\theta}\d\theta\db\theta+G^0_{\phi\phi}\d\phi\db\phi++B_{t\phi}(\d
t\db\phi-\db t\d\phi)\cr
&+{1\over k}\Biggl[(2+F_{11})\d\Phi_1\db\Phi_1+(2+F_{22})\d\Phi_2\db\Phi_2\cr
&\hskip1cm+F_{12}\d\Phi_1\db\Phi_2+
F_{21}\d\Phi_2\db\Phi_1
\cr
&\hskip1cm+(A^1_\phi\d\phi+A^1_t\d t)\db\Phi_1+(A^2_\phi\d\phi+A^2_t\d
t)\db\Phi_2\cr
&\hskip1cm+{\tilde A}^1_t\db t\d\Phi_1
+{\tilde A}^2_t\db t\d\Phi_2,
\Biggr]
\Biggr\}}}
where (for large charge):
\eqn\bosonicsolution{\eqalign{G_{\sigma\sigma}&=1,\,\, G_{\theta\theta}=1,\,\,
G^0_{\phi\phi}=\sin^2\theta+4A^M_\phi A^M_\phi;\cr
G^0_{tt}&=  -{ \cosh\sigma -\delta\over  \cosh\sigma+\delta},\,\,\,
G^0_{t\phi}=   -{ 2\lambda A^M_\phi  \cosh\sigma \over \cosh\sigma+\delta
},\,\,\,
B_{t\phi}=-B_{\phi t} ={2\lambda A^M_\phi \cosh\sigma \over \cosh\sigma+\delta
};\cr
A^1_t&=-4{Q_A \over \cosh\sigma+\delta },\,\,\,
A^2_t=-4{P_A\over \cosh\sigma+\delta};\cr
A^1_\phi&=-8A^M_\phi{\lambda Q_A-Q_B(\cosh\sigma+\delta)\over
\cosh\sigma+\delta},\,\,\,
A^2_\phi=-8A^M_\phi{\lambda P_A-P_B(\cosh\sigma+\delta)\over
\cosh\sigma+\delta};\cr
{\tilde  A}^1_t&=-{4 \cosh\sigma\over k( \cosh\sigma+\delta)},\,\,\,
{\tilde A}^2_t={4\lambda\cosh\sigma\over k(\cosh\sigma+\delta)};\cr
F_{11}&=-8{Q_A \over k(\cosh\sigma+\delta) },\,\,\,
F_{12}=-8{P_A\over k(\cosh\sigma+\delta)};\cr
F_{21}&= 8{\lambda Q_A-Q_B(\cosh\sigma+\delta)\over
k(\cosh\sigma+\delta)},\,\,\,
F_{22}= 8{\lambda P_A-P_B(\cosh\sigma+\delta)\over k(\cosh\sigma+\delta)}.
}}

Notice that the functions $F_{ij}$ and ${\tilde A}^k_t$ appear at subleading
order. After re--fermionisation they become the four--fermi interactions and
tangent space connections for the right movers, respectively.
There is also the dilaton, which may be calculated by various standard methods
(which all amount to the evaluation of the one--loop determinant arising from
the integration over the gauge fields\callan\buscher) to give the action
\eqn\dilatonterm{I_{\rm dilaton}={1\over16\pi}\int d^2z\,\,R^{(2)}{\hat\Phi}}
where $R^{(2)}$ is the world sheet two--dimensional curvature scalar and
\eqn\dilii{{\hat\Phi}=-{1\over2}\log[ \cosh\sigma +\delta]+{\hat\Phi}_0} where
${\hat\Phi}_0$ is a constant.

In the above, the superscript `0' on some of the metric components denotes that
they are   distinct from those   which will appear in the heterotic sigma model
after re--fermionisation, due to large corrections from the $A_\phi,A_t$ gauge
field interactions with the fermions.
As in the 2D cases of preceding sections, the corrections to the metric
components are computed easily in this bosonic form by simply symmetrising and
antisymmetrising the leading $\Phi_1,\Phi_2$ interaction terms, yielding (after
repeated use of the three anomaly equations to simplify expressions):
\eqn\corrections{\eqalign{G_{tt}&=G^0_{tt}-{1\over
4k}\left[(A^1_t)^2+(A^2_t)^2\right]=-{\cosh^2\sigma
-1\over(\cosh\sigma+\delta)^2}\cr
G_{t\phi}&=G^0_{t\phi}-{1\over
4k}\left[A^1_tA^1_\phi+A^2_tA^2_\phi\right]=2\lambda A^M_\phi G_{tt}\cr
G_{\phi\phi}&=G^0_{\phi\phi}-
{1\over 4k}\left[(A^1_\phi)^2+(A^2_\phi)^2\right]=4\lambda^2A^M_\phi A^M_\phi
G_{tt}.
}}
After re--fermionising, the whole model may be written in the standard  fashion
as a heterotic
sigma model:
\eqn\heterotic{\eqalign{I={k\over8\pi}\int d^2z
&\left\{G_{\mu\nu}+B_{\mu\nu}\right\}\d X^\mu\db X^\nu +{k\over8\pi}\int
d^2z\Biggl\{i\lambda^a_R(\db-\Omega_{\mu ab}\db X^\mu)\lambda^b_R\cr&+
i\lambda^\alpha_L(\d-A_{\mu\alpha\beta}\d X^\mu)\lambda^\beta_L
+{1\over2}F_{\mu\nu\alpha\beta}\Psi^\mu_R\Psi^\nu_R
\lambda^\alpha_L\lambda^\beta_L
\Biggr\}}} where the $(a,b)$ and $(\alpha,\beta)$ indices are tangent space and
$U(1)\times U(1)$ current algebra indices respectively.

\newsec{Some 4D spacetime physics.}

The spacetime metric derived from the exact theory \bigaction\ is:
\eqn\anotherform{dS^2={k\over4}
\left\{d\sigma^2-{\cosh^2\sigma-1\over(\cosh\sigma+\delta)^2}
(dt+2\lambda A^M_\phi d\phi)^2+d\theta^2+\sin^2\theta d\phi^2\right\},}
together with the dilaton \dilii\ and  the antisymmetric tensor $B_{t\phi}$ and
 gauge fields $A^i_t,A^i_\phi$ given in \bosonicsolution.  This metric is
invariant under spacetime rotations providing  that $t$ transforms to cancel
the $U(1)$ gauge variation produced by the monopole. This requires $t$ to have
periodicity as $2\lambda\phi$, which is $4\lambda\pi$.  Examining the gauging
\combined, where $U(1)_B$  mixes compact and non--compact variables, this
modding of $t$ can be  anticipated. Another way to see this is to note that
there is a `Dirac string' coordinate singularity in the spacetime which runs
along the $z$-axis ($\theta=0,\pi$). In the Northern hemisphere,
$2A^M_\phi=1-\cos\theta$. The metric is well--behaved everywhere but in the
Southern hemisphere, small loops about the $z$--axis do not shrink to zero
length at the South pole ($\theta=\pi$). So the metric is singular along this
line. This warrants the  use of $2A^M_\phi=-1-\cos\theta$ when down South,
moving the string to the North Pole ($\theta=0$). This change of coordinate
patch (which is the $U(1)$ gauge transformation $A^M\to A^M-1$) is equivalent
to changing variables from $t$ to $t-2\lambda\phi$. Given that $\phi$ has
periodicity $2\pi$, $t$ has periodicity $4\lambda\pi$.
The Dirac string also introduces a quantisation condition on the charges of the
model in the standard way.

This metric is reminiscent of a very special form of the Taub--NUT
metric\ref\taubNUT{E Newman, L Tamburino and T Unti, J Math Phys {\bf 4} (1963)
915.}\ref\misner{C W Misner, J Math Phys {\bf 4} (1963) 924.}, and deserves
further comment. As a solution to the empty space Einstein equations, the full
Taub--NUT metric is:
\eqn\taub{dS^2=f^{-2}dr^2-f^2(dt+2\lambda A^M_\phi
 d\phi)^2+(r^2+\lambda^2)(d\theta^2+\sin^2\theta d\phi^2)} where
\eqn\f{f^2=1-2{Mr+\lambda^2\over r^2+\lambda^2}.}
It reduces to the Schwarzchild metric as $\lambda\to0$. Together with a
constant dilaton, this solution will satisfy the one loop $\beta$--function
equations as a low--energy solution to string theory, as does the Schwarzchild
metric. Given the simple relation between the dilatonic magnetically charged
black hole\gibbons\
\eqn\maggie{\eqalign{dS^2=&\left(1-{2M\over r}\right)^{-1}dr^2
-\left(1-{2M\over r}\right)dt^2
+r^2\left(1-{Q^2\over Mr}\right)(d\theta^2+\sin^2\theta d\phi^2)\cr
{\hat\Phi}-{\hat\Phi}_0=&-{1\over2}\log\left(1-{Q^2\over Mr}\right),}}
and the Schwarzchild solution, it is tempting to conjecture that there exists a
similarly simple modification of the Taub--NUT metric, such that when
supplemented with an antisymmetric tensor with non--zero components
$B_{t\phi}=-B_{\phi t}$, and gauge fields $A_\phi,A_t$, a   dilatonic Taub--NUT
solution to the one loop $\beta$--function equations is constructed.
The solution \maggie\ reduces to the product form at extremality as follows:
First forming the string metric by multiplying the metric \maggie\ by
$\exp2({\hat\Phi}-{\hat\Phi}_0)$, perform the change of variables
$r=2M+\Delta\sinh^2\sigma$ where $\Delta=2M-Q^2/(2M)\to0$ at extremality.
Taking the limit $\Delta\to0$
while holding the dilaton finite in the region of interest by the absorption of
an infinite additive constant into it via
$\exp(-2{\hat\Phi}^\prime_0)=\exp{(-2{\hat\Phi}_0)}\Delta$, yields:
($\tau=t/(2Q)$)
\eqn\extreme{\eqalign{dS^2=&4Q^2\left(d\sigma^2-\tanh^2\sigma
d\tau^2+d\theta^2+\sin^2\theta d\phi^2\right)\cr
{\hat\Phi}-{\hat\Phi}_0^\prime=&-{1\over2}\log\cosh^2\sigma.
}}
This is the low energy solution for which the product of the monopole conformal
field theory of GPS and the supersymmetric linear dilaton black hole conformal
field theory of Witten is the full heterotic string theory\foot{In ref.\GPS\ it
is described how different choices for how the dilaton behaves at extremality
will yield alternatively either the asymptotically flat limit or the mouth
region. The black hole $+$ throat region is the most interesting limit for the
purposes of this paper.}. It is likely that there exists an analogous
transformation to scaled variables for the above conjectured dilatonic
Taub--NUT dyon solution which would yield the low energy solution
\bosonicsolution\ (with the extra background fields \anotherform) as its
extremal limit. The exact string theory is given by the heterotic coset
\bigaction.

This extremal dyonic theory is a natural  generalisation of the simple product
model  \extreme\ which is the familiar extremal magnetically charged black hole
of Gibbons and Maeda\gibbons. It is smoothly connected to it in the limit
$\lambda\to0,\delta\to1$.  The gauge group of this model is $U(1)\times U(1)$,
as it is  necessary to have $P_A$ and $Q_B$ distinct from $Q_A$ and $P_B$
(respectively), in order to preserve the smooth limit to the $U(1)$ magnetic
case as $\lambda\to0$ while cancelling all of the anomalies to all orders, as
discussed in section 6. Sending $P_A$ and $Q_B$ to zero in this limit recovers
   the $U(1)$ gauge group.

Because of the lack of a $\sigma$--dependent factor multiplying the round
2--sphere metric, the solution does not have any behaviour  resembling   the
approach to an asymptotically  flat limit\foot{The full Taub--NUT is
asymptotically flat, although it is not possible to write the flat limit in
global Cartesian form $G_{\mu\nu}\sim\eta_{\mu\nu}+O(1/\sigma)$.}. This and all
of the solutions studied in this paper as simple   products of the angular and
radial theories of sections (3) and (4) become throat or wormhole--like in the
large $\sigma$ limit.  This is the familiar   throat geometry  typical of
extremal black hole solutions. The radius of the throat is set by the   charges
of the model as can be seen from \anotherform\ and \simpleranomaly. It is   a
simple matter to evaluate the electric  and magnetic charges on this extremal
black hole solution. They are given by integrating    the electromagnetic field
strengths $F^i$ and their  duals ${}^*F^i$
 over  the asymptotic 2-sphere at infinity:
\eqn\charge{\eqalign{Q^i_E=&{1\over4\pi}\int_{S^2}
{}^*F^i_{\theta\phi}(\sigma\to\infty)\cr
Q^i_M=&{1\over4\pi}\int_{S^2}F^i_{\theta\phi}(\sigma\to\infty).}} where a
dilaton factor is necessary to take the dual\foot{This may be deduced by
studying the action for the effective field theory  in the standard
way\callan.}.
This yields
\eqn\charges{Q^1_E=4Q_A,\,Q^2_E=4P_A,\,Q^1_M=4Q_B,\,Q^2_M=4P_B.}

These four charges are completely independent as the construction began with
eight free parameters, from which four degrees of freedom were removed by the
three anomaly equations \cancel\ together with the condition on the central
charge of the theory \central.
There is also a dilaton charge and two type of axion charge on the theory.
These however are determined in terms of $Q^i_E$ and $Q^i_M$, given that there
were only four free parameters left over from the construction of the model.
As there is a Dirac string, the usual quantisation conditions for $U(1)$
magnetic and electric charges apply, requiring the product of the electric and
magnetic charges to be integer (in units of $\hbar c/2$).

\newsec{Conclusions and Outlook.}

This paper has presented methods for constructing a rich class of exact
heterotic string backgrounds using what might be called a `heterotic coset'
technique\foot{This is in contrast to earlier constructions\ref\jimroger{S J
Gates Jr and W Siegel, Phys Lett {\bf 206} (1988) 631\semi S Bellucci, R Brooks
and J Sonnenschein, Mod Phys Lett {\bf A3} (1988) 1537.}\ of heterotic string
backgrounds using   ungauged $(0,1)$ supersymmetric WZW models  via superfield
constructions.}. This technique combines a gauged WZW model with supersymmetric
right--moving fermions and current--algebra left--moving fermions in a way
which produces a conformally invariant theory. The novelty here is that none of
the three   ingredients above need be conformally invariant on its own. This
introduces a close relationship between the charges of the fermions and the
level of the WZW model. The role of bosonisation and  subsequent
re--fermionisation is emphasised as crucial in determining  the correct gauge
invariant lagrangian and in the later determination of the correct spacetime
metric of the heterotic string background. These features will be common to any
model constructed using this type of mixing of fermions and gauged WZW
models\foot{In ref.\ref\rabinovici{A Giveon, E Rabinovici and A A Tseytlin,
Nucl Phys {\bf 409} (1993) 339. hep-th/9304155}, the use of supersymmetric
gauged WZW models truncated to $(0,1)$ supersymmetry was considered as a means
of constructing heterotic string backgrounds. It was noted there that in order
to cancel the 2d gauge anomaly, the addition of  internal fermions was an
option. They were coupled to an external (spacetime) gauge field. However, the
only external gauge fields  which were considered in that paper were such that
the theory was again effectively $(1,1)$ supersymmetric.  This is in contrast
to the more general couplings considered in the present paper, allowing truly
$(0,1)$ (or $(0,2)$) supersymmetric models to be defined. See also
ref.\ref\tseytlin{K Sfetsos and A A Tseytlin, Nucl Phys {\bf  B415} (1994) 116.
hep-th/9308018.} for a discussion of heterotic string backgrounds using `chiral
gauged' WZW models. The author is grateful to A A Tseytlin for directing his
attention to refs.\rabinovici\tseytlin.}.

The application considered in this paper was the construction of dyonic
extremal charged black holes. The prototype of the angular theory of the models
is the `monopole' conformal field theory of Giddings, Polchinski and
Strominger\GPS, which was presented by them as an orbifold of an $SU(2)$ WZW
model. Here it is shown to be equivalent to a heterotic coset of the type
presented in this paper. Similar constructions applied to $SL(2,\rline)$
yields a family of 2D charged black hole solutions. These are exact conformal
field theories  which reduce to the one--loop solutions of McGuigan, Nappi and
Yost\chiara, and a special case of the bosonic solution of Ishibashi, Li and
Steif\ishibashi.

Products of these with the monopole theory will give 4D models which may be
interpreted as 4D dyonic extremal black holes. Also considered was a model
where the radial and angular sectors were mixed by embedding $U(1)\times U(1)$
non--trivially into $SL(2,\rline)\times SU(2)$. This yielded a 4D dyonic model
with a metric resembling what  would result from extremising a conjectured
dilatonic Taub--NUT metric.  Such a dilatonic Taub--NUT (with torsion) would be
interesting  for many reasons. The Taub--NUT solution has interesting
cosmological interpretations\misner\ and the existence of a stringy Taub--NUT
would be interesting in that context. Also, (Euclidean) Taub--NUT  is a
self--dual gravitational instanton\ref\hawking{S W Hawking, Phys Lett {\bf 60A}
(1977) 81.}, and so a string theoretic solution in such a background is
certainly interesting and deserves further study. An axionic, dyonic and
extremal stringy Taub--NUT with wormhole, found here as an exact conformal
field theory may be a useful solution for studying string  instantons
representing spacetime topology change\ref\strom{S B Giddings and A Strominger,
Nucl Phys {\bf B306} (1988) 890.}.
The use of Euclidean Taub--NUT in constructing monopole solutions to
Kaluza--Klein theories as first presented by  Sorkin\ref\sorkin{R D Sorkin,
Phys Rev Lett, {\bf 51} (1983) 87.}, and Gross and Perry\ref\gross{D J Gross
and M J Perry, Nucl Phys {\bf B226} (1983) 29.} may also be relevant. The
presence of such solutions in string theory is important\ref\banks{T Banks, M
Dine, H Dijkstra and W Fischler, Phys Lett {\bf B212} (1988) 45.}. It would be
easy to construct that type of solution (at least in some  limit) using these
methods. The compact time in the solution of this paper would become another
internal coordinate while the role of time would be introduced by tensoring the
heterotic coset with a free field.

The theories constructed in this paper all have $(0,2)$ world sheet
supersymmetry. The consequences of this were not pursued in this paper,
although the pure magnetic theory was shown in ref.\GPS\ to not satisfy the
conditions for $N=1$ spacetime supersymmetry.
Leaving behind the black--hole applications, the construction of this paper can
certainly be applied to the study of $(0,2)$ string compactification. These are
less well--studied than their more specialised $(2,2)$ cousins, but are
extremely important for realisations of string vacua which may be relevant to
the world in which we live.

\bigskip
\bigskip

{\bf Acknowledgments.}

I am grateful to Ed Witten for awakening my interest in 4D stringy black holes
and for his   encouragement and guidance throughout the project. Discussions at
various stages of this project with Paul Aspinwall, Per Berglund, Martin
Bucher, Curt Callan, M\aa ns Henningson, Juan Maldacena and Kate Okikiolu were
also useful. Part of this work was supported by a Lindemann Fellowship and an
SERC Fellowship.

\bigskip
\bigskip

\centerline{\bf Note Added.}

\bigskip

After this paper was submitted for publication, the paper of ref.\ref\andy{D A
Lowe and A Strominger, {\sl `Exact Four--Dimensional Dyonic Black Hole and
Bertotti--Robinson Spacetimes in String Theory'}, preprint UCSBTH--94--14,
hep-th/9403186.}\ appeared, in which 4D dyonic black hole solutions and
Bertotti--Robinson Spacetimes were constructed as conformal fields theories,
using analogous techniques to those employed in ref.\GPS. In particular, an
$SL(2,\rline)/Z$ orbifold is central to the construction of those solutions.
Those models are  a special case of the models described in section~4. This is
easily seen by a choice of gauge analogous to that used in section~3 to show
the equivalence between the GPS $SU(2)$ orbifold and the gauged WZW model
described there.

\listrefs
\bye